\documentclass[prx, reprint]{revtex4-1}

\usepackage{graphicx}
\usepackage{amsmath}
\usepackage{amssymb}
\usepackage{url}
\usepackage{bm}
\usepackage{xcolor}

\begin{document}

\title{Unidirectional light transport in dynamically modulated waveguides} 



\author{Momchil Minkov}

\author{Shanhui Fan}
\email[]{shanhui@stanford.edu}
\affiliation{Department of Electrical Engineering, and Ginzton Laboratory, Stanford University, Stanford, CA 94305, USA}

\date{\today}

\begin{abstract}
One-way edge states at the surface of photonic topological insulators are of significant interest for communications, nonlinear and quantum optics. Moreover, when reciprocity is broken in a photonic topological insulator, these states provide protection against disorder, which is of particular importance for slow light applications. Achieving such a one-way edge state, however, requires the construction of a two-dimensional structure. Here, we show how unidiriectional Floquet bands can arise in purely one-dimensional, adiabatically-modulated dynamic systems, in contrasts with the higher dimensionality needed in topological insulators. We also show that, using realistic experimental parameters, the concept can be implemented  using both a coupled-resonator optical waveguide and a photonic crystal waveguide. Furthermore, we illustrate the associated protection against disorder, and find it to be of a novel nature when compared to Floquet topological insulators. 
\end{abstract}

\maketitle

\section{Introduction}

Photonic devices incorporating slow light can serve to enhance optical nonlinearities and light-matter interactions, and as optical delay lines for information storage \cite{Krauss2008, Baba2008}. The group index $n_g = c/v_g$, defined as the ratio of the speed of light in vacuum to the group velocity in a slow-light device, is thus a particularly important figure of merit \cite{Schulz2010}. Using periodic structures like a coupled-resonator optical waveguide (CROW) \cite{Yariv1999} or a photonic crystal (PhC) waveguide \cite{Baba2008}, the group index can in fact be made arbitrarily large at the frequency at the edge of the photonic Brillouin zone -- at least in theory. In practice, however, small fabrication imperfections introduce strong scattering of the photonic modes into modes propagating in the opposite direction \cite{Patterson2009, Mazoyer2009, Minkov2013}, leading to a degradation of the transport, and, in the extreme case, to Anderson localization of light \cite{John1987}. For a fixed disorder magnitude, these undesirable effects grow stronger with increasing $n_g$, inevitably setting a limit on the highest value achievable in experiment.

Recently, topological photonics has become a strong research focus \cite{Lu2014}, motivated both by an interest in the fundamental physical properties of topological insulators, and by the promise of unidirectional, disorder-immune propagation of light \cite{Raghu2008, Haldane2008, Wang2009, Hafezi2011, Rechtsman2013, Fang2012, Minkov2016, Bahari2017}, which could lift the limit on the group index that is imposed by disorder in slow-light structures. However, the unidirectional frequency bands of topological insulators always correspond to states on the edges of a two- (or higher-) dimensional system \cite{Lu2014}. This is required for the non-trivial winding of the one-dimensional edge bands in the Brillouin zone, but it literally adds an extra dimension of complexity to the fabrication of waveguides based on this effect. Recently, it was realized that the \textit{Floquet} quasi-frequency bands that arise in periodically-modulated systems \cite{Shirley1965} have a topological classification that is richer than that of ordinary frequency bands \cite{Kitagawa2010}, which arises from their periodicity in quasi-energy (or quasi-frequency) space. One illustration of this is the appearance of `anomalous' edge states in gaps between quasi-frequency bands with a zero associated Chern number difference \cite{Rudner2013}, but this effect arises once again in a two-dimensional structure. The only previous discussion of a unidirectional band in a one-dimensional system was given in Ref. \cite{Kitagawa2010}, but only through an idealized model with no corresponding physical realization.  

In this paper we show that, within the adiabatic approximation, unidirectional Floquet quasi-frequency bands can be achieved in a purely one-dimensional, dynamically-modulated waveguide. This leads to a significant simplification of the structures needed for unidirectional light transport. Specifically, we show how such bands can be achieved both in a modulated CROW, and in a modulated PhC waveguide using experimental parameters relevant to state-of-the-art integrated photonic devices. Furthermore, we find that many of the appealing properties of the edge states in topological insulators are preserved, most importantly -- robustness with respect to imperfections in the system. We also discuss the details of this disorder protection, and highlight that it is different, and in some cases superior, to that of photonic topological insulators based on dynamic modulation \cite{Fang2012, Minkov2016}.

The paper is organized as follows. In Section \ref{sec:floq_ad} we provide the theoretical background. In Section \ref{sec:ccw}, we illustrate how a unidirectional Floquet band with a constant group velocity can be implemented in a modulated CROW, and show the associated robustness to disorder, as compared with a standard slow-light CROW. In Section \ref{sec:phc}, we extend these results to the case of a modulated photonic crystal waveguide. Finally, in Section \ref{sec:conclusion}, we discuss some experimental considerations regarding the implementation of our proposed devices and the group index that can be expected, as well as the nature and magnitude of the protection against disorder. 

\section{Floquet bands and adiabatic evolution}
\label{sec:floq_ad}

We start with an overview of Floquet theory for time-periodic systems \cite{Shirley1965}. Consider a quantum mechanical Hamiltonian $\hat{H}(t)$ such that $\hat{H}(t) = \hat{H}(t + T)$ for a given period corresponding to a modulation frequency $\Omega = 2\pi/T$. For such systems, the Floquet theorem can be employed, asserting that the evolution of any state under $\hat{H}(t)$ can be written as a linear combination of Floquet quasi-eigenstates defined as
\begin{equation}
|\phi_{\alpha}(t)\rangle = e^{-i\epsilon_{\alpha} t} |v_{\alpha}(t)\rangle, 
\end{equation}
where $\epsilon_{\alpha}$ are the quasi-energies, $\alpha$ is an eigenmode index, and the states $|v_{\alpha}(t)\rangle$ are time-periodic with period $T$ and determined by the eigenvalue equation 
\begin{equation}
(\hat{H}(t) - i\partial_t)|v_{\alpha}(t)\rangle = \epsilon_{\alpha} |v_{\alpha}(t)\rangle.
\label{eqn:Floq_ham} 
\end{equation}
The states $|\phi_{\alpha}(0)\rangle$ are themselves eigenstates of the time-evolution operator $\hat{U}(t) = \mathcal{T}\exp\left(-i\int_0^t \hat{H}(t')\mathrm{d}t'\right)$ at time $t = T$, with eigenvalues determined by the quasi-energies as
\begin{equation}
\hat{U}(T) |\phi_{\alpha}(0)\rangle = |\phi_{\alpha}(T)\rangle = e^{-i\epsilon_{\alpha} T} |\phi_{\alpha}(0)\rangle. \label{eqn:floq_U}
\end{equation}
We note that the quasi-energies are only defined modulo $\Omega$, i.e. the time-periodicity introduces periodicity in frequency-space. 

\begin{figure}
\centering
\includegraphics[width = 0.47\textwidth, trim = 0in 0in 0in 0in, clip = true]{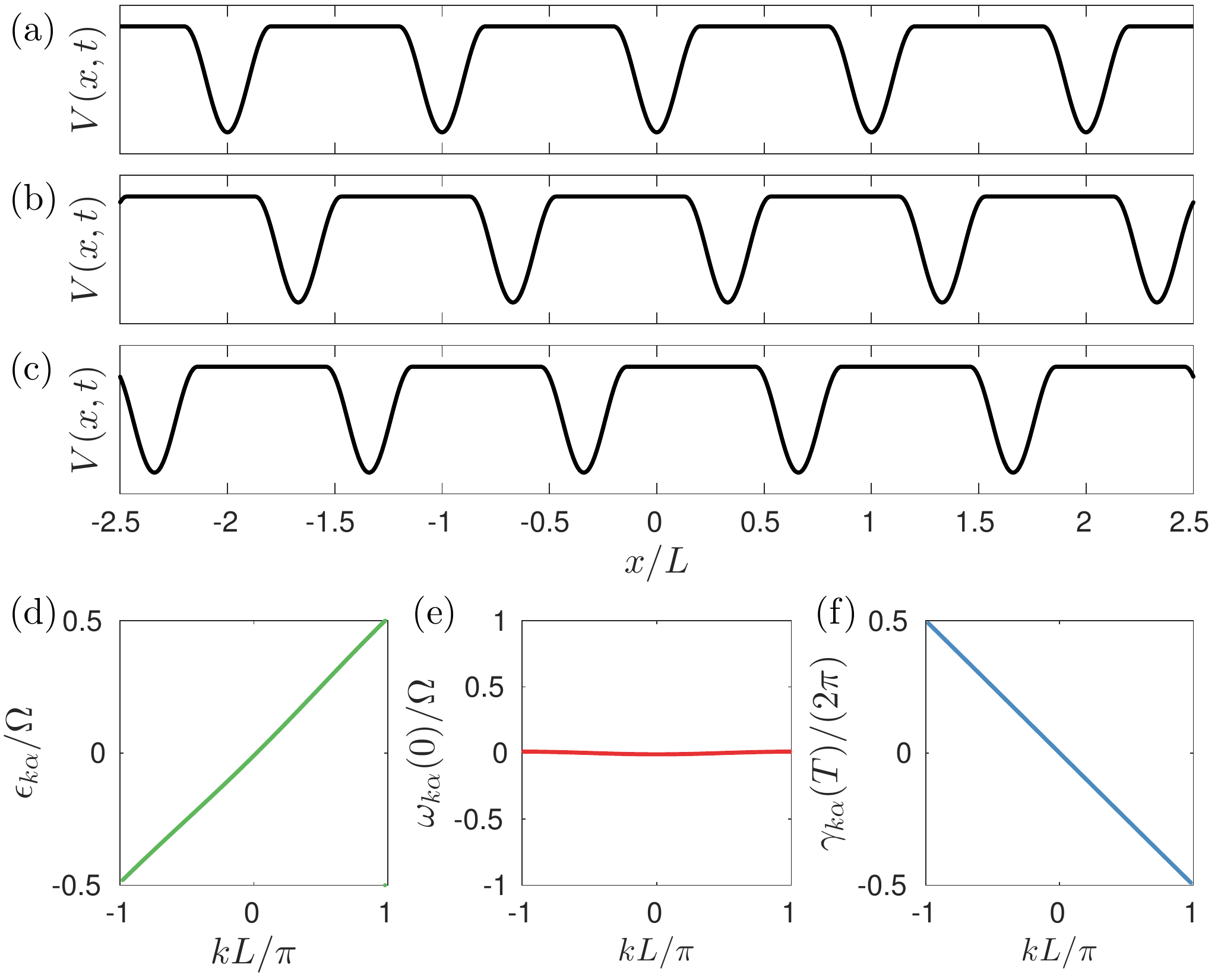}
 \caption{(a)-(c): Schematic of a lattice of potential wells uniformly sliding to the left, shown at times (a): $t = 0$; (b): $t = T/3$; and (c): $t = 2T/3$. (d)-(f): In the limit of deep wells, (d): quasi-energy band; (e): instantaneous frequency band at $t = 0$; and (f): Berry phase after one period.} 
\label{fig_floq}
\end{figure}

As a simple physical example, which is also related to both systems that we study later on in this paper, consider the Hamiltonian $\hat{H} = V(x, t)|x\rangle \langle x|$ corresponding to a potential that is uniformly sliding towards the positive-$x$ direction, $V(x, t) = V(x - v t, 0)$. If $V(x, 0)$ also has spatial periodicity with period $L$, then the potential is time-periodic with period $T = L/v$. A schematic example is illustrated in Fig. \ref{fig_floq}(a)-(c), where we show a periodic lattice of potential wells separated at a distance $L$, uniformly sliding to the right. The spatial periodicity of $V(x, t)$, which is preserved at all times, also means that the Bloch momentum $k$ is conserved (modulo $2\pi/L$).

Ref. \cite{Kitagawa2010} put forth an intuitive derivation of the quasi-energy band $\varepsilon_{k}$ corresponding to a sliding potential as the one in Fig. \ref{fig_floq}(a)-(c), in the limit in which the wells are sufficiently deep such that the dynamics can be projected on the basis consisting of only the localized states. We repeat this here for pedagogical purposes. In this limit, a starting state $|\psi(0)\rangle = |\psi_x \rangle$ localized at position $x$ moves together with its potential well, such that $|\psi(t)\rangle = |\psi_{x - vt}\rangle$. Defining the reciprocal-space states $|\psi_k \rangle$ such that $|\psi_k (t)\rangle = \sum_x e^{-ikx} |\psi_{x-vt}\rangle$, we then find $|\psi_k(T) \rangle = \sum_x e^{-ikx} |\psi_{x-L} \rangle = \sum_x e^{-ik(x + L)} |\psi_{x} \rangle = e^{-ikL}|\psi_k(0)\rangle$. Thus, for every $k$, $|\psi_k(t)\rangle$ is a Floquet quasi-eigenstate as per eq. (\ref{eqn:floq_U}), with an associated quasi-energy $\varepsilon_{k} = kL/T$. This is illustrated in Fig. \ref{fig_floq}(d), and it can be seen that it has a non-trivial winding that is only possible because of the folding of the Brillouin zone in quasi-energy space. 

The significance of such a non-trivial winding of the Floquet band -- in particular for photonic systems -- comes from the fact that the Floquet quasi-energy of a band defines a time-averaged group velocity of a wavepacket in the same way as the band frequencies of a static system determine the group velocity. Namely, assume we have a starting wavepacket $\varphi(x, t = 0)$, expanded on the basis of the Bloch wavefunctions $|\varphi_{k\alpha}(t)\rangle = e^{ikx}|u_{k\alpha}(t)\rangle$:
\begin{equation}
\varphi(x, t = 0) = \int_{\mathrm{BZ}} \mathrm{d}k \mathcal{W}(k) e^{ikx} |u_{k\alpha}(0)\rangle.
\end{equation}
We assume further that the expansion coefficients $\mathcal{W}(k)$ are narrowly centered around some Bloch vector $k_0$, and Taylor-expand
\begin{align}
\epsilon_{k\alpha} &= \epsilon_{k_0 \alpha} + (k - k_0)\left. \frac{\partial \epsilon_{k\alpha}}{\partial k} \right|_{k_0} \\ &\equiv \epsilon_{k_0 \alpha} + (k - k_0)\bar{v}_\alpha (k_0). \nonumber
\end{align}
Using eq. (\ref{eqn:floq_U}), we find at $t = T$
\begin{align}
\varphi(x, T) = &e^{ik_0 x - i\epsilon_{k_0 \alpha} T} \times \\ \nonumber &\int_{\mathrm{BZ}} \mathrm{d}k \mathcal{W}(k) e^{i(k-k_0)(x - \bar{v}_\alpha (k_0) T)}|u_{k\alpha}(0)\rangle.
\end{align}
The position of the wavepacket thus shifts by $\bar{v}_\alpha(k_0)T$ after every period $T$, which justifies interpreting $\bar{v}_\alpha(k_0)$ as the group velocity of band $\alpha$ at $k_0$, averaged over one cycle. The intuitive relationship between $\bar{v}$ and the Thouless charge transport \cite{Thouless1983} has been discussed in Ref. \cite{Kitagawa2010}, where it was shown that the integral of the Floquet group velocity $\bar{v}_\alpha (k)$ over the Brillouin zone is equal to the charge pumped over one cycle associated to the filled band $\alpha$, as discussed by Thouless. However, as opposed to solid-state systems, filled bands do not naturally arise in photonic systems. The group velocity, on the other hand, is still a very important figure of merit, determining for example the maximum delay as well as the strength of the light-matter interaction in delay lines. Furthermore, a constant, unidirectional group velocity in the entire Brillouin zone implies complete absence of backscattering and hence robustness in the presence of disorder introduced into the waveguide. Thus, the significance of the Floquet band of Fig. \ref{fig_floq}(d) goes beyond Thouless pumping. 

The main result of this Section is now to derive an expression for the Floquet quasi-energies that is generally valid in the adiabatic limit, but does not assume an infinitely deep potential. This will allow us to study realistic systems, and to propose in Sections \ref{sec:ccw} and \ref{sec:phc} physical photonic structures in which a Floquet dispersion like the one of Fig. \ref{fig_floq}(d) can be implemented. We start from the \textit{instantaneous} eigenstates of the Hamiltonian $\hat{H}(t)$, defined as 
\begin{equation}
\omega_{k\alpha}(t) |\psi_{k\alpha}(t)\rangle = \hat{H}(t)|\psi_{k\alpha}(t)\rangle, 
\label{eqn:inst_eig}
\end{equation}
where $\omega_{k\alpha}(t)$ denote the instantaneous eigen-frequencies. These states form a complete basis set at every time $t$, and thus the time evolution of any arbitrary state in the system can be expanded as
\begin{equation}
 |\Psi_k(t)\rangle = \sum_n a_{k\alpha}(t) |\psi_{k\alpha}(t)\rangle e^{-i\theta_{k\alpha}(t)}, 
 \end{equation} 
where $\theta_{k\alpha}(t) = \int_0^t \omega_{k\alpha}(t') \mathrm{d}t'$. The Schr\"{o}dinger equation can thus be re-written as a system of coupled differential equations for the expansion coefficients $a_{k\alpha}$, namely
 \begin{equation}
\frac{\partial}{\partial t} a_{k\beta} = - \sum_\alpha a_{k\alpha} \left\langle \psi_{k\beta} \left| \frac{\partial}{\partial t} \right| \psi_{k\alpha} \right\rangle e^{i(\theta_{k\beta}(t) - \theta_{k\alpha}(t))}	
\label{eqn:ad_th1}
\end{equation}
This is thus far an exact result. Now, if we assume a slowly-varying Hamiltonian, to first order in the time derivative the solution expanded around a starting instantaneous eigen-state $|\psi_{k\alpha}\rangle$ is given by \cite{Sakurai1994, Thouless1983, Berry1984, Xiao2010} 
\begin{align}
|\Psi_{k\alpha}(t)\rangle =& e^{-i\theta_{k\alpha}(t)} e^{i\gamma_{k\alpha}(t)} \times \label{eqn:psi_time} \\ &\left(|\psi_{k\alpha}(t)\rangle - i \sum_{\beta \neq \alpha} K_{\alpha\beta}(k, t)|\psi_{k\beta} (t) \rangle \right), \nonumber
\\
K_{\alpha \beta}(k, t) =& \frac{\langle \psi_{k\beta} (t)| \partial/\partial t |\psi_{k\alpha} (t)\rangle}{\omega_{k\alpha}(t) - \omega_{k\beta}(t)}, \label{eqn:K}
\end{align}
valid in the limit $|K_{\alpha \beta}(k, t)| \ll 1$, $\forall \, t, \beta$. For subsequent use we refer to $K_{\alpha\beta}(k,t)$ as the \textit{overlap factor}. Its magniitude measures how well the adiabatic condition is satisfied. The perfect adiabatic evolution is achieved when the overlap factor approaches zero. The quantity $\gamma_{k\alpha}(t)$ is the Berry phase, i.e. the integral over the Berry connection, for band momentum $k$ and band $\alpha$: 
\begin{equation}
\gamma_{k\alpha}(t) = i\int_0^t \langle \psi_{k\alpha} (t')| \partial/\partial t |\psi_{k\alpha} (t')\rangle \mathrm{d}t'
\end{equation}
Since the eigenstates of eq. (\ref{eqn:inst_eig}) are only defined up to an arbitrary gauge that depends on $k$ and $t$, this quantity is generally also gauge-dependent. However, in the case of cyclic dynamics, $\hat{H}(t + T) = \hat{H}(t)$, the Berry phase $\gamma_{k\alpha}(T)$ (modulo $2\pi$) becomes a gauge-independent, observable quantity \cite{Berry1984, Xiao2010}. 
Now, assuming adiabatic evolution, as well as single-valued instantaneous eigenstates such that $|\psi_{k\alpha}(0)\rangle = |\psi_{k\alpha}(T)\rangle $, the state $|\Psi_{k\alpha}(t)\rangle$ as given in eq. (\ref{eqn:psi_time}) obeys 
\begin{equation}
|\Psi_{k\alpha}(T)\rangle = e^{-i\theta_{k\alpha}(T)}e^{i\gamma_{k\alpha}(T)} |\Psi_{k\alpha}(0)\rangle,
\end{equation}
and is a Floquet quasi-eigenstate with quasi-energy
\begin{equation}
\epsilon_{k\alpha}T = \theta_{k\alpha}(T) - \gamma_{k\alpha}(T).
\end{equation}
For the case of the uniformly sliding potential of Fig. \ref{fig_floq}(a)-(c), the instantaneous frequencies $\omega_{k\alpha}(t)$ are in fact time-independent. More generally, assuming that $\omega_{k\alpha}(t)$ is approximately time-independent on the scale of $\Omega$, as will be the case in our subsequent examples, we obtain simply
\begin{equation}
\epsilon_{k\alpha}/\Omega = \omega_{k\alpha}(0)/\Omega - \gamma_{k\alpha}(T)/(2\pi). \label{eqn:eps_om}
\end{equation}

This relationship is illustrated schematically in Fig. \ref{fig_floq}(d)-(f) for the case of the deep sliding potential. The instantaneous eigen-frequency band $\omega_{k\alpha}$ is flat in this limit. Thus, the winding of the quasi-energy band implies that the Berry phase also has to wind in the Brillouin zone, as shown in Fig. \ref{fig_floq}(f). In other words, we could infer the Berry phase based on our knowledge of the quasi-energies. However, more generally, and in realistic systems, the utility of eq. (\ref{eqn:eps_om}) goes the other way around. Namely, there is a variety of ways to compute the instantaneous eigenstates of eq. (\ref{eqn:inst_eig}), and thus both terms on the right-hand side of eq. (\ref{eqn:eps_om}). These can then be used to compute the Floquet dispersion when the evolution is adiabatic, which is particularly useful in practice, since direct numerical and analytic calculations of the quasi-energy band by diagonalization of eq. (\ref{eqn:Floq_ham}) can be far more involved as compared to the computation of the instantaneous band structure. 

The main conclusion of this Section is that a sliding potential can lead to unidirectional waveguiding that is expected to be robust to disorder. Based on Fig. \ref{fig_floq} and our discussion above, we can more specifically identify three requirements needed for the unidirectional Floquet band: (1) we need a Berry phase that winds in the Brillouin zone, (2) we need adiabatic evolution, and (3) we need an instantaneous starting band that is narrow-band when compared to the modulation frequency. In the next Section, we show how these conditions can be met in realistic physical systems, and how the unidirectional band emerges as predicted.

\section{Implementation in a generic coupled-cavity waveguide}

\label{sec:ccw}

\subsection{Setup and theory}

\begin{figure}
\centering
\includegraphics[width = 0.47\textwidth, trim = 0in 0in 0in 0in, clip = true]{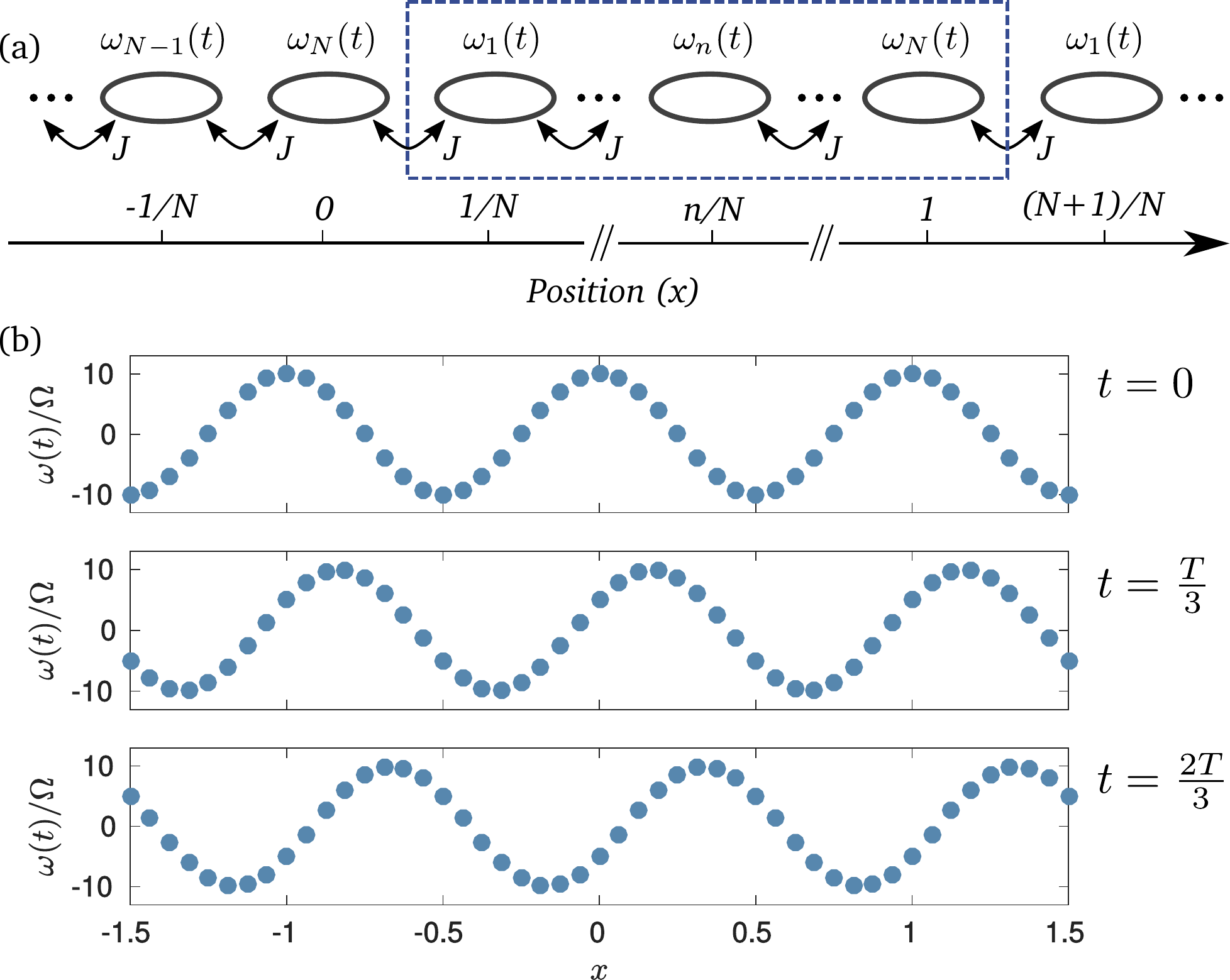}
 \caption{(a): Schematic of the system. There are $N$ cavities within an elementary cell of unit length (blue dashed rectangle), with first-neighbor coupling $J$ and time-varying resonance frequencies $\omega_n(t)$. (b): Resonance frequency vs. position for $A/\Omega = 10$, $N = 16$, at times $t = 0$, $T/3$, and $2T/3$.}
\label{fig_ccw}
\end{figure}

The first system that we consider is a spatially-discrete analogue of a uniformly sliding cosine potential, $V(x, t) = \cos(2\pi x/L - \Omega t)$. Namely, we study a CROW as in Fig. \ref{fig_ccw}(a), in which the resonance frequency of each cavity $\omega_i(t)$ is sinusoidally modulated in time. We set the starting, unmodulated frequency of each cavity to $\omega_0 = 0$, since a non-zero $\omega_0$ would only appear as a constant frequency offset in all the results presented below. We further impose a real-space periodicity $N$, such that $\omega_i(t) = \omega_{i+N}(t)$, and denote the cavity positions $x_m = m/N$ for integer $m$, such that the unit cell is of unit length. The coupled-mode theory equations describing this system \cite{Haus1984, Fan2003, Minkov2017} can be written in the second-quantization form of a single-particle Hamiltonian as \cite{Fang2012}
\begin{align}
H(t) = \sum_{m} A \cos(2\pi x_{m} - \Omega t) a_{m}^{\dagger}a_{m} + \\ \nonumber
(J a_{m}^{\dagger} a_{m+1} + h.c.),
\end{align}
where the operator $a^{\dagger}_{m}$ creates a particle at position $x_{m}$. Fig. \ref{fig_ccw}(b) shows the spatial distribution of $\omega_i(t)$ for $A/\Omega = 10$, $N = 16$, at three instants, $t = 0$, $t = T/3$, $t = 2T/3$. Defining the $k$-space operators
\begin{equation}
a^{\dagger}_{kn} = \sum_p e^{ikx_{pN + n}} a^{\dagger}_{pN + n}, \quad \quad n = 0\dots N-1, \quad p \in \mathcal{Z},
\end{equation}
the Hamiltonian becomes
\begin{equation}
H(t) = \int_{-\pi}^\pi \mathcal{A}_k^\dagger \mathcal{H}(k, t) \mathcal{A}_k \mathrm{d}k,
\end{equation}
with $\mathcal{A}^{\dagger}_k = (a_{k1}^\dagger, a_{k2}^\dagger, \dots a_{kN}^\dagger )$, and 
\begin{align}
&\mathcal{H}(k, t) \label{eqn:Hk} = \\ \nonumber &\begin{pmatrix}
\omega_1(t) & J e^{ik/N} & 0 & \dots & 0 & J e^{-ik/N} \\
Je^{-ik/N} & \omega_2(t) & Je^{ik/N} & \dots & 0 & 0 \\
\vdots & \vdots & \vdots & \vdots & \vdots & \vdots  \\
0 & 0 & 0 & \dots & \omega_{N-1}(t)  & Je^{ik/N} \\
Je^{ik/N} & 0 & 0 & \dots & J e^{-ik/N} & \omega_N(t) 
\end{pmatrix}.
\end{align}
Diagonalizing the matrix $\mathcal{H}(k, t)$ thus yields the instantaneous eigen-frequencies $\omega_{k\alpha}(t)$.
We can also numerically compute the quasi-energies using standard Floquet theory. The states $|v_{k\alpha}(t)\rangle$ of eq. (\ref{eqn:Floq_ham}) are space- and time-periodic, and can be expanded on the basis $|n, p\rangle_k = e^{ip\Omega t} a_{kn}^\dagger |0\rangle$, where $p$ is an integer, i.e. 
\begin{equation}
 |v_{k\alpha}(t)\rangle = \sum_{n, p} v_{k\alpha}(n, p) |n, p\rangle_k.
\end{equation}
The inner product defining the Hilbert space of the Floquet Hamiltonian $(H(t) - i\partial_t)$ of eq. (\ref{eqn:Floq_ham}) is defined as $\langle\langle \bullet | \bullet \rangle \rangle = \frac{1}{T}\int_0^T \langle \bullet | \bullet \rangle \mathrm{d}t$. We can thus compute the non-zero matrix elements of the Floquet Hamiltonian in the  $|n, p\rangle_k$ basis as
\begin{align}
  \label{eqn:Floq_mat}
& \langle _k\langle n', p' | H - i\partial_t | n, p \rangle_k\rangle = \\ \nonumber &=\begin{cases}
p\Omega & \quad p = p', n = n' \\
\mathcal{H}_{nn'}(k, 0) & \quad p = p', n \neq n' \\
(A e^{i2\pi n/N})/2 & \quad p = p'-1, n = n' \\
(A e^{-i2\pi n/N})/2 & \quad p = p'+1, n = n'
  \end{cases} 
\end{align}
where $\mathcal{H}_{nn'}$ denote the matrix elements of eq. (\ref{eqn:Hk}). The Floquet Hamiltonian can thus be readily diagonalized numerically by restricting the matrix elements of eq. (\ref{eqn:Floq_mat}) to a certain order $p_{\max}$ such that $|p|, |p'| < p_{\max}$. This value is chosen high enough to achieve convergence.

\begin{figure}
\centering
\includegraphics[width = 0.5\textwidth, trim = 0in 0in 0in -0.0in, clip = true]{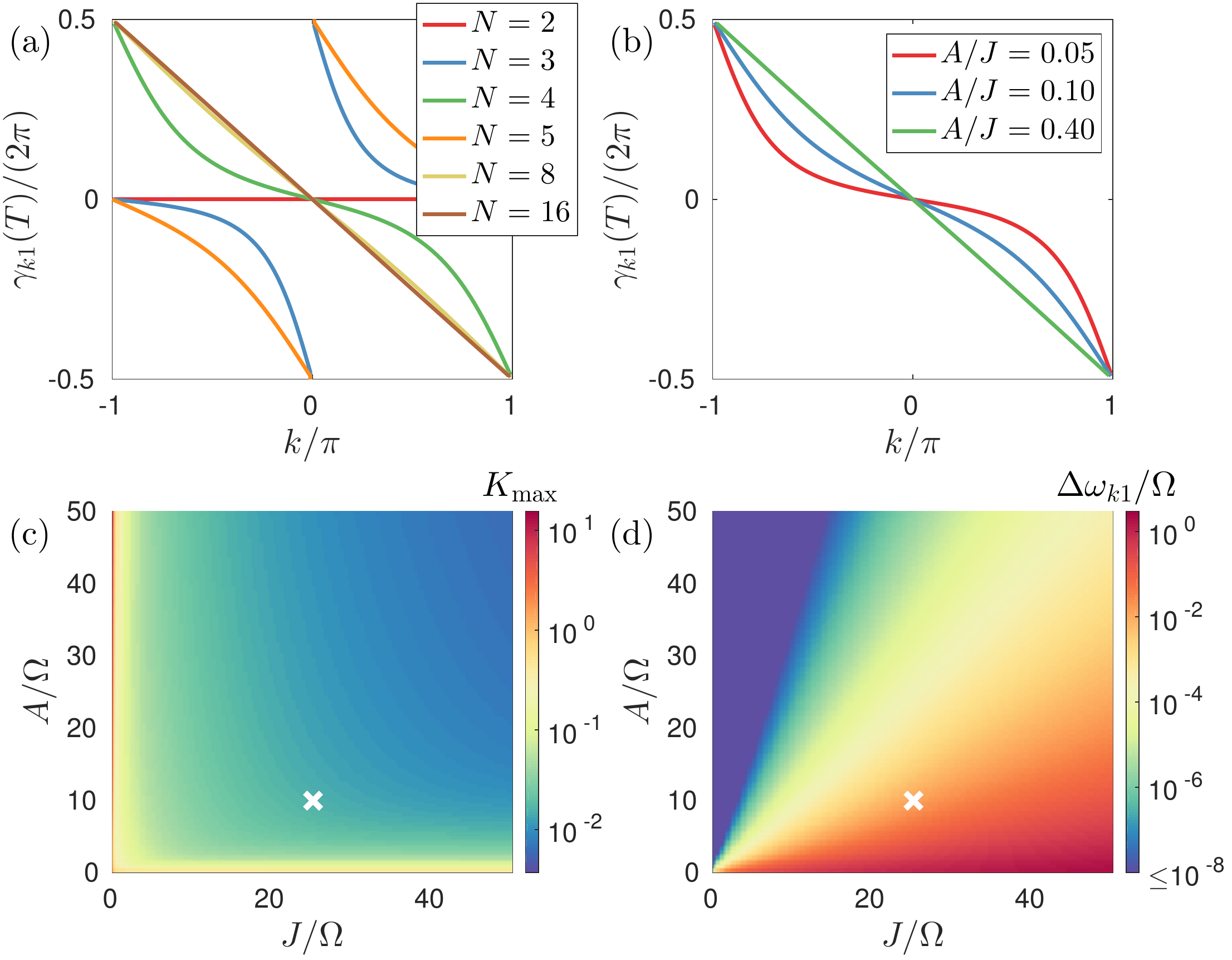}
 \caption{(a)-(b): Berry phase of the lowest-frequency band for a chain with (a): $A/J = 1$ and several values of $N$, and (b): $N = 16$ and several values of $A/J$. (c): Maximum overlap term $K_\mathrm{max}$ (see text) as a function of $A$ and $J$, for $N = 16$. (d): Same, but for the bandwidth of the lowest-frequency band. The white cross marks the parameters used in subsequent Figures in this Section.}
\label{fig_ccw_berry}
\end{figure}

We can now explore the dependence on the system parameters of some of the quantities relevant to our target structure. In Fig. \ref{fig_ccw_berry}(a)-(b), we show the dependence of the Berry phase of the lowest-frequency band of the CROW for various values of $N$ and $A/J$. This was obtained numerically by computing the instantaneous eigenstates on a discretized mesh in time. As can be seen, the Berry phase winds around the Brillouin zone for all values of $N$ apart from $N = 2$, in which case the system is time-reversal invariant. With increasing $N$ and $A$, the Berry phase approaches the simple dependence $\gamma_{k1}(T) = -k$, as can be expected from our previous discussion of an infinitely deep, continuous potential. 

As discussed in Section \ref{sec:floq_ad}, a winding Berry phase is one of three conditions needed for achieving a fully unidirectional Floquet band. The other two are adiabatic evolution, as well as a narrow starting band $\omega_{k\alpha}(0)$. Thus, in Fig. \ref{fig_ccw_berry}(c), we plot the maximum magnitude of the overlap term for the first band as defined in eq. (\ref{eqn:K}), i.e. $K_{\mathrm{max}} = \mathrm{max}_{k, t, \beta}(|K_{1\beta}(k, t)|)$, as a function of $A$ and $J$, with $N = 16$. For the adiabatic condition to hold, we need $K_\mathrm{max} \ll 1$, which in turn requires both $J > \Omega$ and $A > \Omega$. Intuitively, when $J$ goes to zero, the states become fixed at individual lattice sites and cannot follow the moving potential, while when $A$ goes to zero, the potential becomes too shallow and the states are not bound to the local minima. In Fig. \ref{fig_ccw_berry}(d), we plot the bandwidth $\Delta \omega_{k1}$ of the lowest-frequency instantaneous band at $t = 0$, which increases with $J$ and decreases with $A$. Still, as can be seen, there is a broad range of possible parameters that fit all requirements. For the remainder of this Section, we set $N =16$, $A/\Omega = 10$, and $J/\Omega = 25$ (white crosses in Fig. \ref{fig_ccw_berry}(c)-(d)), for which $K_\mathrm{max} = 0.015 \ll 1$ and $\Delta \omega_{k1}/\Omega = 0.009 \ll 1$.

\subsection{Floquet bands}
\label{sec:ccw_bands}

\begin{figure}
\centering
\includegraphics[width = 0.48\textwidth, trim = 0in 0in 0in -0.0in, clip = true]{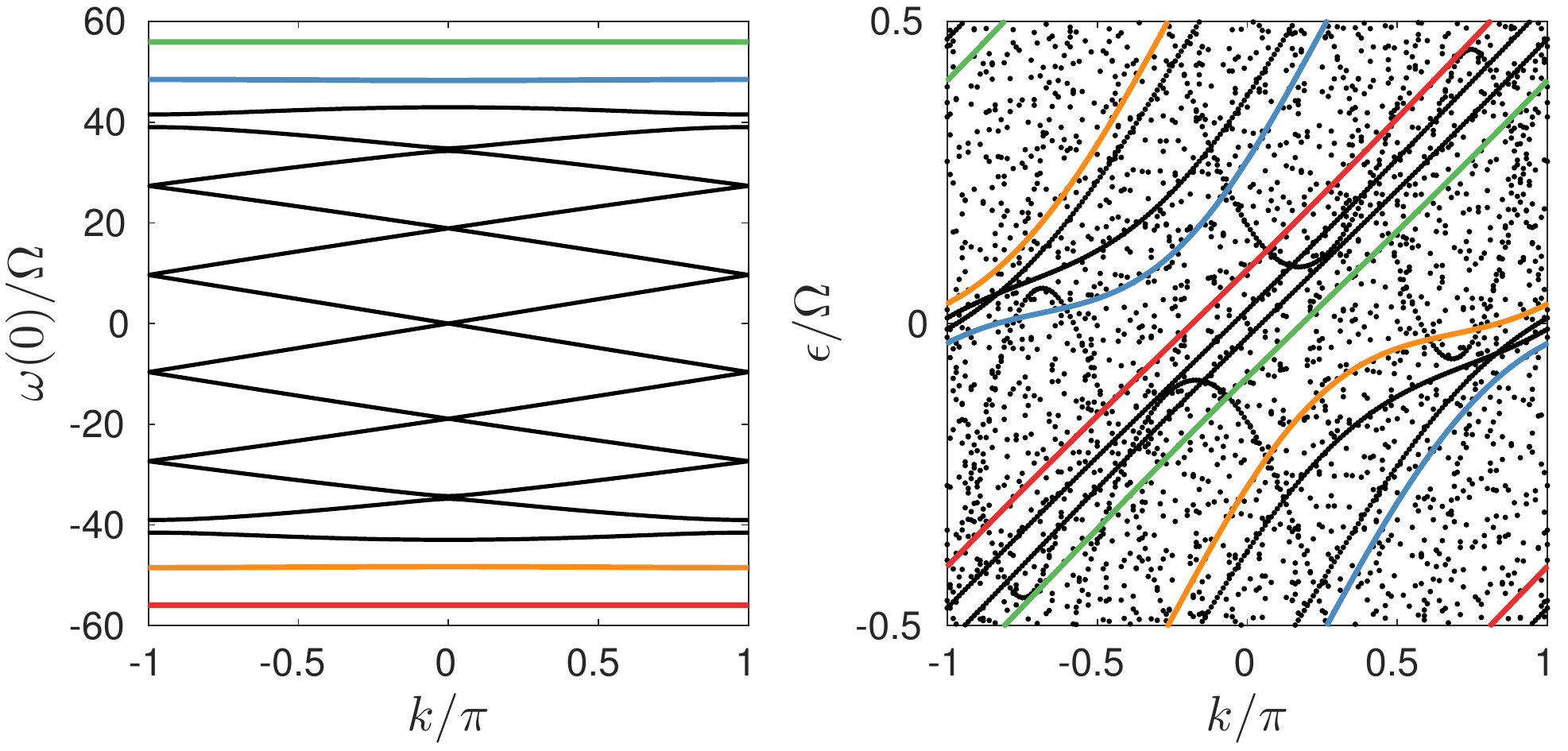}
 \caption{For a chain with $A/\Omega = 10$, $J/\Omega = 25$, $N = 16$, (a): instantaneous band-structure at $t = 0$, and (b): Floquet quasi-energy bands. Black dots are computed through exact diagonalization. The colored lines are computed from eq. (\ref{eqn:eps_om}) for the correspondingly-colored bands in (a).} 
\label{fig_ccw_floquet}
\end{figure}

We can now test the main result of Section \ref{sec:floq_ad} using the concrete physical system as described above. In Fig. \ref{fig_ccw_floquet}(a), we plot the instantaneous band structure $\omega_{k\alpha}(0)$ of the CROW, while in Fig. \ref{fig_ccw_floquet}(b), we show with black dots the exact Floquet bands computed numerically after diagonalizing eq. (\ref{eqn:Floq_mat}). We note once again that the quasi-energy axis is folded with period $\Omega$. This, together with the fact that some of the bands in panel (a) have a bandwidth significantly larger than $\Omega$, results in the speckled appearance of the exact quasi-energies, as the bands are re-folded many times into the frequency Brillouin zone. Importantly, however, several continuous black bands stand out, each of which has a bandwidth on the scale of $\Omega$. We will now show through eq. (\ref{eqn:eps_om}) that these bands can be associated with adiabatically guided states from the two lowest- and highest-frequency bands of panel (a). 

We first look at the lowest-frequency band, plotted in red in panel Fig. \ref{fig_ccw_floquet}(a). For this band, we compute and plot in panel (b) the band of the quasi-energy using eq. (\ref{eqn:eps_om}), with the associated Berry phase taken from the green curve in Fig. Fig. \ref{fig_ccw_berry}(b). The resulting line matches well one of the black bands from the exact diagonalization, apart from a small systematic offset. Next, we note that there is a symmetry of the bands with respect to $\omega = 0$, and so for example between the red and the blue bands in panel (a). In particular, the frequency in such a pair of symmetric bands is given by $\omega_{k1}(0) = -\omega_{k16}(0)$, while we also find numerically that the term $K_\mathrm{max}$ and the Berry phase $\gamma_{k\alpha}(T)$ are the same. Thus, the green line in Fig. \ref{fig_ccw_floquet}(b) shows the adiabatic prediction associated to the highest-frequency band in panel (a), and it accounts for the second straight band visible in black. We can repeat the same procedure for bands number two and fifteen (orange and blue lines in Fig. \ref{fig_ccw_floquet}(a), respectively), for which the term $K_\mathrm{max} = 0.028$ is still much smaller than one, justifying the application of eq. (\ref{eqn:eps_om}). The corresponding results are again shown in Fig. \ref{fig_ccw_floquet}(b), and account very well for the remaining two continuous black bands visible in the panel. The offset here between the analytic calculation and the exact numerical diagonalization arises since the adiabatic condition is not strictly satisfied, i.e $K_{\mathrm{max}}$ is not strictly zero. We have checked, using different parameters $A$, $J$, and $N$, that the difference between the analytic prediction of eq. (\ref{eqn:eps_om}) and the exact quasi-frequencies decreases with decreasing $K_\mathrm{max}$. 

In short, as seen in Fig. \ref{fig_ccw_floquet}(b), we have achieved fully unidirectional Floquet bands, which are also well accounted for by eq. (\ref{eqn:eps_om}).

\subsection{Unidirectional source emission}

\begin{figure}
\centering
\includegraphics[width = 0.5\textwidth, trim = 0in 0in 0in 0in, clip = true]{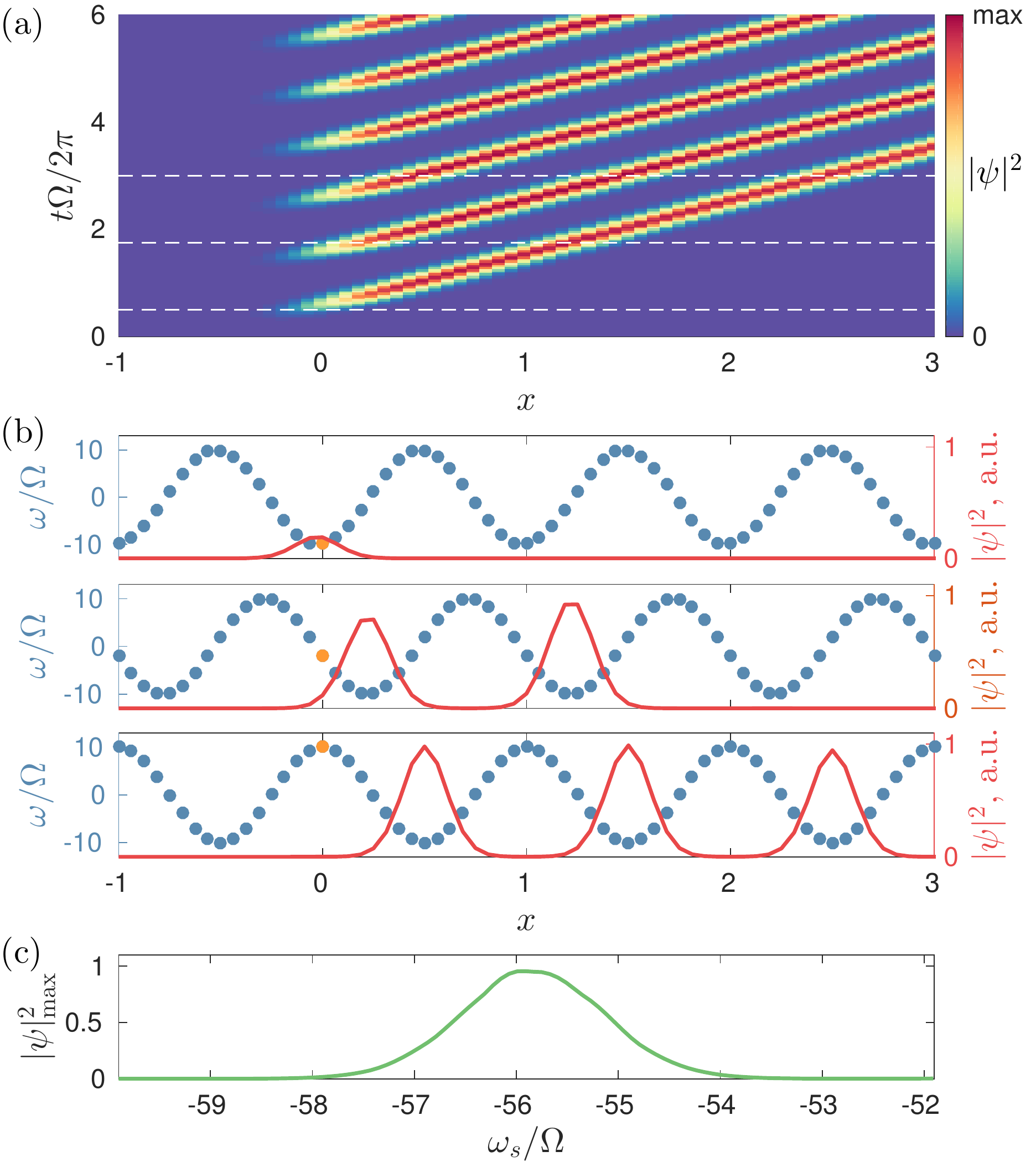}
 \caption{(a): Time evolution of the field intensity in the CROW with $A/\Omega = 10$,  $J/\Omega = 25$, $N = 16$, given a source of frequency $\omega_s/\Omega = -55.9$ in the cavity at $x = 0$. (b): Resonance frequencies and field intensity at times $t = 0.5T$, $t = 1.75T$, and $t = 3T$ (shown with white dashed lines in (a)). The $x = 0$ cavity where the source is located is marked in orange. (c): Peak of the emitted pulses versus source frequency.} 
\label{fig_source}
\end{figure}

As shown in Section \ref{sec:ccw_bands}, the dynamically modulated CROW can exhibit a unidirectional quasi-energy band. In order to use this band to demonstrate unidirectional light transport, we will need to selectively excite it. Moreover, as light propagates, a mode in such a unidirectional band should not couple to other modes of the system. In general, in a Floquet system, any two quasi-energy bands that intersect can couple to one another. Thus, in principle, a mode from a given unidirectional band of Fig. \ref{fig_ccw_floquet}(b) would couple with any other mode in the system that has the same quasi-energy and wavevector. In our case, however, due to the adiabatic consideration, an excitation of a mode  in the lowest band of the instantaneous band structure is expected to stay in the same band. Consequently, the coupling of modes in the quasi-energy band formed from the lowest instantaneous band to all the other modes in the system should be minimal, in spite of the fact that the quasi-energy bands form a near continuum due to the folding along the quasi-energy axis, as shown in Fig. \ref{fig_ccw_floquet}(b). Therefore, to demonstrate unidirectional light transport, it is sufficient to place an excitation source in one of the resonators, and choose the frequency of the excitation source to be close to the frequency of the lowest band of the instantaneous band structure. 

In Fig. \ref{fig_source}(a), we show a dynamic simulation of the field intensity $|\psi(x, t)|^2$ inside the CROW of Fig. \ref{fig_ccw_floquet}, assuming a continuous-wave source at frequency $\omega_s/\Omega = -55.9$ (i.e. close to $\omega_{k1}$) placed in the cavity at position $x = 0$. In panel (b), we show snap-shots of $\omega(x, t)$ and $|\psi|^2$ at three different times (white dashed lines in panel (a)). As can be seen, the source emits the strongest at the times when the cavity in which it is placed is at its lowest frequency. This is because the instantaneous eigenstates corresponding to $\omega_{k1}$ are, at any given time, localized around the lowest-frequency cavity region. The instantaneous eigenstates in fact look very similar to the emitted pulses shown in Fig. \ref{fig_source}. 

As expected due to the linear, one-way Floquet band, the result of the dynamic simulation looks qualitatively the same regardless of the source frequency, as long as it only couples with the first band, i.e. $|\omega_s - \omega_{k1}| \ll |\omega_s - \omega_{k,i\neq 1}|$. The difference, however, is the maximum intensity of the emitted pulses, which are the strongest when the source is exactly resonant with $\omega_{k1}$. This is shown in Fig. \ref{fig_source}(c), where we plot the maximum intensity vs. source frequency. We note that the bandwidth of the resonance is comparable to $\Omega$, and significantly larger than that of the instantaneous band $\omega_{k1}$, which is much smaller than $\Omega$. 

\subsection{Disorder-protected delay line}

\begin{figure}
\centering
\includegraphics[width = 0.47\textwidth, trim = 0in 0in 0in 0in, clip = true]{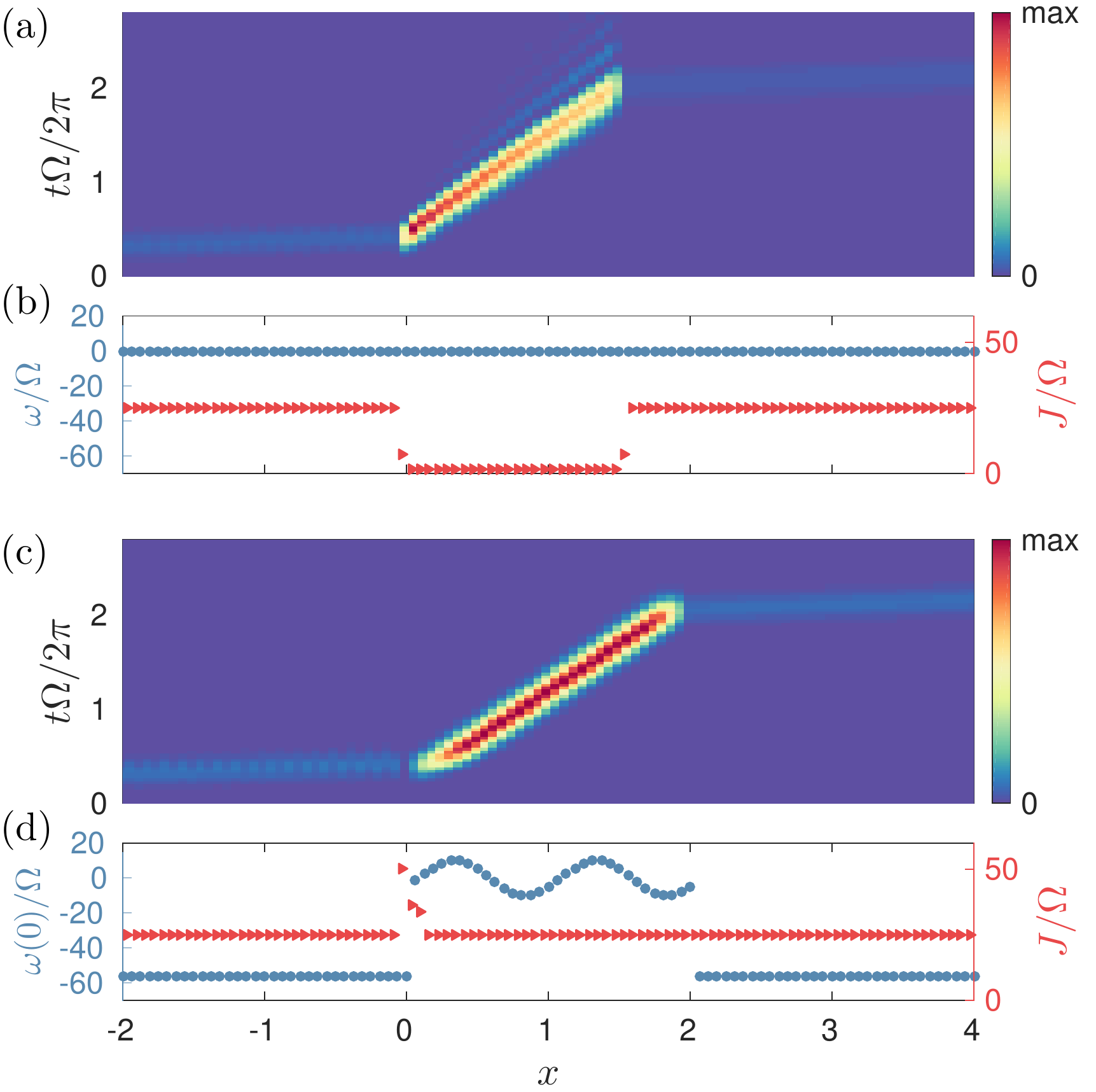}
 \caption{(a): Time evolution of the field intensity in- and out-coupled to a classical delay line. There are $N = 16$ cavities within a unit length on the $x$-axis. (b): All cavities have $\omega = 0$ at all times, while the coupling constant is $J_f/\Omega = 25$ in the fast-light region $x < 0$ and $x > 1.4$, and $J_s/\Omega = 8/(2\pi)$ in the slow-light region $x \in [0, 1.4]$. (c)-(d): Field intensity, resonance frequencies at time $t = 0$, and coupling constants for a delay line based on the modulated-CROW concept.} 
\label{fig_delay}
\end{figure}

Next, we show how a unidirectinal optical delay line protected against disorder can be built on the basis of the modulated CROW. In Fig. \ref{fig_delay}(a)-(b), we first show a regular delay line made from an unmodulated CROW. Specifically, we consider a chain of cavities with a nominal $N = 16$ (i.e. 16 resonators per unit length), but with a resonance frequency $\omega_0 = 0$ at all times for all cavities. The dispersion is then given by $\omega(k) = 2J\cos(k/N)$, the group velocity of a pulse centered around $\omega_0$ is correspondingly $v_g(k = \pi/2) = 2J/N$, and it can thus be controlled through the coupling constant $J$. In Fig. \ref{fig_delay}(a)-(b), we plot a dynamic simulation of a pulse propagating through a fast-light region with $J_f/\Omega = 25$ for cavities at $x < 0$, which then enters a slow-light region with $J_s/\Omega = 8/(2\pi)$, such that the group velocity is $v_g = 1/T$, i.e. one elementary cell in time $T$. The slow-light region extends to $x = 1.4$, at which point the fast-light value $J_f$ is introduced again. The starting pulse is $\psi(x, t = 0) = e^{ik_0x}e^{-(x-x_0)^2/(2\sigma_x^2)}$, with $k_0 = \pi/2$, $x_0 = -8$, and $\sigma_x = 2.5$, corresponding to a bandwidth of $\Delta \omega/\Omega = 1.25$. To ensure strong in-coupling of the pulse to the slow-light region, we apodize the coupling constants at the interfaces between the fast- and the slow-light regions \cite{Sumetsky2003}. In particular, we minimize the reflection of the pulse with respect to the coupling constants between the cavities at position $-2/N$ and $-1/N$, $-1/N$ and $0$, and $0$ and $1/N$, which results in $J_a/\Omega = [24.9, 7.5, 1.8]$, respectively in the apodized region (Fig. \ref{fig_delay}(b)). For the out-coupling of the pulse, these apodized coupling constants are taken in reverse order. As can be seen in  Fig. \ref{fig_delay}(a), the pulse enters completely the slow-light region with minimum reflection (intensity reflection coefficient $< 1\%$), and is indeed slowed down in the slow-light region. We note that the visible broadening in that region is due to group-velocity dispersion effects, since the pulse bandwidth is now comparable to the total bandwidth $4J_s$ of the slow-light region.

In Fig. \ref{fig_delay}(c)-(d) we plot the same pulse, but this time using the modulated CROW as a delay line. More precisely, in the spatial region $x \in [0, 2]$ we take the modulated chain of Figs. \ref{fig_ccw_floquet} and \ref{fig_source}, with $N = 16$, $A/\Omega = 10$, $J/\Omega = 25$ (Fig \ref{fig_delay}(d)). The modulated CROW thus has the same group velocity as the unmodulated slow-light CROW of panel (a). However, light in- and out-couples faster, which is why we use a slightly longer slow-light region to achieve the same delay. In the regions $x < 0$ and $x > 2$, we have the same unmodulated fast-light CROW as in panel (a), only this time we set the resonance frequency of every cavity to $\omega_0/\Omega = -55.9$, such that the central pulse frequency is resonant with the lowest instantaneous band $\omega_{1k}$ of the modulated structure. We note that in this setup, the in-coupling of the pulse depends on its starting position, or, alternatively, on the starting time $t_0$ of the modulation. Furthermore, to ensure strong transmission, we again need to apodize the coupling constants $J_a$, defined above. Thus, we numerically minimize the reflection at the interface with respect to $J_a$ and $t_0$, which leads to $J_a/\Omega = [50.0, 36.1, 33.7]$, $t_0 = 0.34T$, with a corresponding reflection of less than $3\%$. Notably, however, the couplings at the output end do not need to be apodized: there are no available back-propagating states for the pulse there, and it out-couples with unity transmission. Finally, we note that the starting pulse bandwidth was chosen such that it matches the one of the outgoing pulse, which is approximately given by $\Omega$. With these design, the modulated CROW indeed achieves the same group velocity reduction as the unmoduated CROW, with a similarly high overall transmission coefficient for the pulse used. 

\begin{figure}
\centering
\includegraphics[width = 0.47\textwidth, trim = 0in 0in 0in 0in, clip = true]{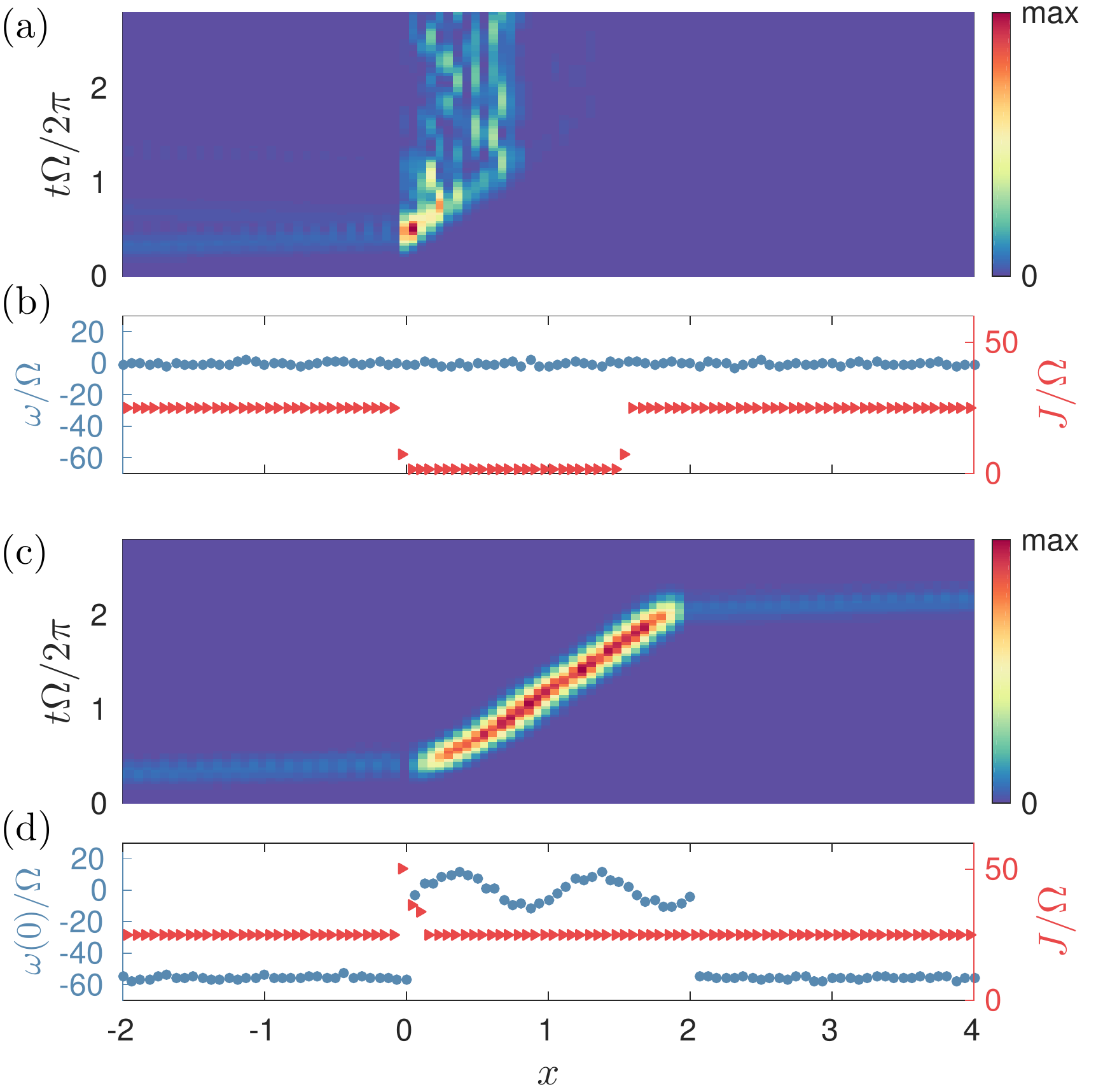}
 \caption{Same as Fig. \ref{fig_delay}, but including random Gaussian disorder in $\omega$ with a standard deviation $\sigma/\Omega = 1$.} 
\label{fig_delay_dis}
\end{figure}

The superiority of our modulated scheme over a standard delay line is illustrated in Fig. \ref{fig_delay_dis}, where we plot the same pulse propagation as in Fig. \ref{fig_delay}, but now assuming random Gaussian disorder in the resonance frequency of each cavity with zero mean and standard deviation $\sigma = \Omega$. Disorder of this type is very common in real systems, and, in regular CROWs, its detrimental effect grows stronger as the group velocity decreases. Indeed, as can be seen in Fig. \ref{fig_delay_dis}(a), in the fast-light region the pulse is not strongly affected by the disorder, since $J_f \gg \sigma$. However, inside the slow-light region we have $J_s \approx \sigma$. This leads to more reflection at the interface between the two domains, while the light that does enter the slow-light region is strongly distorted and localized. In sharp contrast, the propagation through the modulated CROW, shown in Fig. \ref{fig_delay_dis}(c)-(d), works in the same way as in the disorder-less case of Fig. \ref{fig_delay}(c)-(d). This is due to the fact that here $A \gg \sigma$ and $J \gg \sigma$ everywhere, and thus the disorder effects are much weaker. The numerical results demonstrate that the unidirectional quasi-energy band structure can indeed be used to overcome disorder-induced backscattering in CROW structures. 

\section{Implementation in a Photonic Crystal waveguide}
\label{sec:phc}

\subsection{Setup and theory}

We will now show how the ideas developed in the previous Sections can also be implemented in a photonic crystal waveguide, opening our results to a broader class of integrated photonic devices. First, we derive the dynamics of electromagnetic radiation in a photonic structure under a time-dependent permittivity modulation, using an approach similar to that of Section \ref{sec:floq_ad}, but starting from the Maxwell's equations of the system. This can be done in the spirit of the seminal works on topological photonics \cite{Haldane2008, Raghu2008}, in which the Berry phase associated to electromagnetic modes has been defined. We assume no free charges and currents, relative magnetic permeability $\mu = 1$ everywhere, and an isotropic, lossless material with an instantaneous dielectric response such that the permittivity $\varepsilon(\mathbf{r}, t)$ is real, scalar, and does not depend on the frequency $\omega$. At any fixed $t$, Maxwell's equations can then be written as a generalized Hermitian eigenvalue problem for the instantaneous eigenmodes $\bm{u}_\mu = (\mathbf{H}_\mu, \mathbf{E}_\mu)^T$, with $\mathbf{E}$ and $\mathbf{H}$ the electric and the magnetic fields, respectively:
\begin{align}
\begin{pmatrix}
 0 & -i\bm{\nabla} \times \\
 i\bm{\nabla} \times & 0 
\end{pmatrix}
\bm{u}_\mu(t) = 
\omega_\mu(t) \begin{pmatrix}
\mu_0 \mathbf{I} & 0 \\
0 &  \varepsilon_0 \varepsilon(\mathbf{r}, t) \mathbf{I}
\end{pmatrix} \bm{u}_\mu,
\label{eqn:umu}
\end{align}
where $\mathbf{I}$ is the $3\times 3$ identity matrix. We can thus, as in Section \ref{sec:floq_ad}, expand the general dynamics of the electromagnetic system on the basis of these orthonormal instantaneous eigenmodes, such that 
\begin{align}
&\bm{u}(t) = \sum_\mu c_\mu(t) \bm{u}_\mu(t) e^{i\theta_\mu(t)},
\end{align}
with $\theta_\mu(t) = -\int_0^t \omega_{\mu}(t') \mathrm{d}t'$. Using eq. (\ref{eqn:umu}), Maxwell's equations can then be re-written as coupled differential equations for the expansion coefficients:
\begin{align}
&\dot{c}_\nu  = - \sum_\mu c_\mu (\bm{u}_\nu, \dot{\bm{u}}_\mu) e^{i(\theta_\mu - \theta_\nu)}, \label{eqn:u}
\end{align}
where time-derivative is denoted by a dot, and the inner product is defined as 
\begin{equation}
(\bm{u}_\nu, \dot{\bm{u}}_\mu) = \int \mathrm{d}\mathbf{r} \bm{u}_\nu^\dagger \begin{pmatrix}
\mu_0 \mathbf{I} & 0 \\
0 &  \varepsilon_0 \varepsilon(\mathbf{r}, t) \mathbf{I} 
\end{pmatrix} \dot{\bm{u}}_\mu
\end{equation}
The discussion is thus far exact. In analogy with Section \ref{sec:floq_ad}, we can similarly define the overlap factor whose magnitude defines adiabacity,
\begin{align}
 K_{\mu\nu}(t) &= \frac{(\bm{u}_{\nu}, \dot{\bm{u}}_{\mu})}{\omega_{\mu} - \omega_{\nu}},
 \label{eqn:phc_K}
 \end{align}
 as well as the Berry phase acquired under adiabatic evolution,
\begin{align}
 \gamma_{\nu}(t) &= i\int_0^t (\bm{u}_\nu(t'), \dot{\bm{u}}_\nu(t')) \mathrm{d}t', \label{eqn:phc_berry}
\end{align}

\begin{figure}
\centering
\includegraphics[width = 0.47\textwidth, trim = 0in 0in 0in 0in, clip = true]{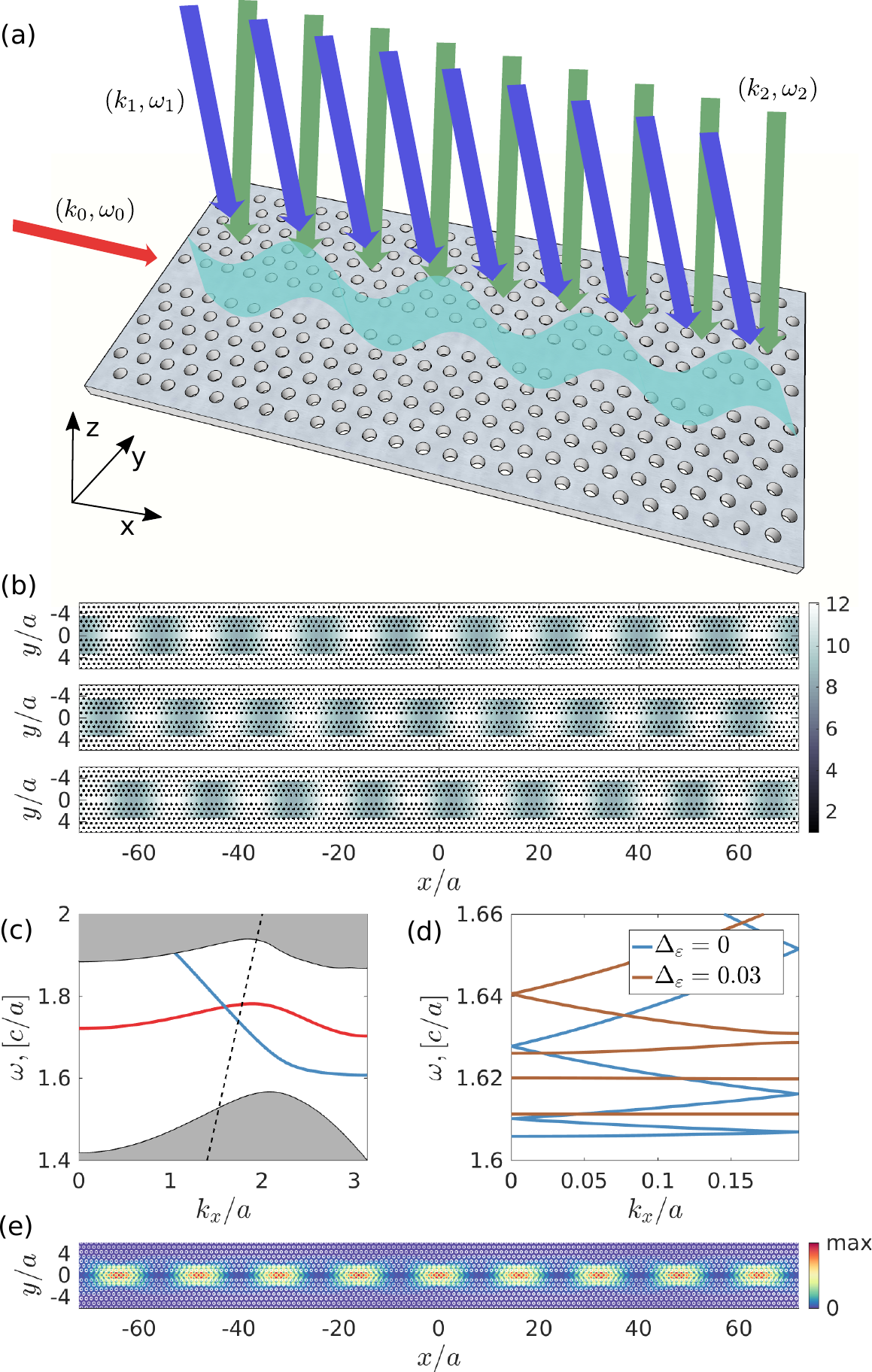}
 \caption{(a): Schematic of the simulated setup -- a photonic crystal waveguide is illuminated from above by two interfering fields, which generate a traveling-wave intensity pattern. (b): Permittivity $\varepsilon(\mathbf{r}, t)$ of the modulated structure with $l_x = 16a$ and $\Delta_\varepsilon = 0.3$, at $z = 0$ and times $t = 0$, $t = T/3$, and $t = 2T/3$ . (c): Photonic bands of the unmodulated PhC. The two guided bands (red and blue) have opposite symmetry with respect to the $xz$-plane. The black dashed line is the light cone. (d): Brown: band structure of the modulated waveguide at $t = 0$, with $l_x = 16a$ and $\Delta_\varepsilon = 0.03$. Blue: band structure of the unmodulated waveguide computed with a supercell of length $l_x$ in the $x$-direction. (e): Electric field $|\mathbf{E}_{k_x}(\mathbf{r})|^2$ of the lowest-frequency guided mode corresponding to the brown bands in panel (c), with $k_x = 0$.} 
\label{fig_schematic}
\end{figure}

We now apply this formalism to a model photonic crystal system, shown schematically in Fig. \ref{fig_schematic}(a). We consider a silicon-slab PhC W1 waveguide, formed in a triangular lattice of circular holes, with one missing row of holes. The physical parameters are, relative permittivity $\varepsilon_{S} = 12.1$ in silicon and $\varepsilon_1 = 1$ in air, slab thickness $d = 220$nm, lattice constant $a = 400$nm, and hole radius $r = 100$nm. The eigenfrequencies and the full electromagnetic eigenmodes of this structure can be efficiently and reliably simulated using the guided-mode expansion method \cite{Andreani2006, Minkov2015}. In particular, we use a computational cell of length $L_x$ in the $x$-direction and $L_y$ in the $y$-direction, while the $z$-direction is included analytically. Everywhere below, we set $L_y = 7\sqrt{3}a$, while $L_x$ is determined by the periodicity in the $x$-direction. The Bloch momentum $k_x$ is a conserved quantity, with a Brillouin zone of width $2\pi/L_x$. The Bloch bands of the unmodulated PhC ($L_x = a$) are shown in Fig. \ref{fig_schematic}(c). The physical parameters were chosen such that the operational frequency $\omega_0$, roughly defined by the slow-light region of the lowest guided band (blue), corresponds to a free-space wavelength close to $1550$nm.

To achieve modulation, we assume that the structure has an intensity-dependent index, and is illuminated from above with two plane waves at slightly different frequencies $\omega_1, \omega_2$, lying above the photonic band-gap of the underlying PhC lattice. Furthermore, we assume the plane waves have the wavevectors $\mathbf{k}_1 = (k_1, 0, \sqrt{\omega_1^2/c^2 - k_1^2})$, $\mathbf{k}_2 = (k_2, 0, \sqrt{\omega_2^2/c^2 - k_2^2})$, corresponding to waves with slightly different angles of incidence (Fig. \ref{fig_schematic}(a)). We further assume that the difference in the $k_z$-components is small enough, such that the electric field of the combined beam is approximately constant in the $z$-direction on the length-scale of the slab thickness $d$ (this is justified for the range of parameters we use below). Thus, the electric field intensity inside the slab due to the beating of these two illuminating plane waves can be written as 
\begin{equation}
I(x, y, t) = I_0\times(2 + 2\cos(\Delta_k x - \Omega t)),
\end{equation}
with $\Delta_k = k_2 - k_1$ and $\Omega = \omega_2 - \omega_1$. Then, assuming an optical material non-linearity that gives rise to an intensity-dependent refractive index, this leads to an optically-induced modulation of the refractive index. For concreteness, we consider free-carrier dispersion, which has already been used in a similar setup \cite{Leonard2002, Beggs2012}. For simplicity, we assume that the response is instantaneous with respect to $\Omega$, which in practice sets an upper bound on $\Omega$ that depends on the details of the implementation. The permittivity in the material then becomes $\varepsilon_S \rightarrow \varepsilon_S + \Delta \varepsilon(\mathbf{r}, t)$, with a permittivity change due to the illuminating waves given by
\begin{equation}
\Delta \varepsilon(\mathbf{r}, t) = - f I_0 \times (2 + 2\cos(\Delta_k x - \Omega t)),
\end{equation}
We rewrite Eq. (29) as
\begin{equation}
\Delta \varepsilon(\mathbf{r}, t) = -
\varepsilon_{S}\Delta_\varepsilon \sin^2 \left(\pi\frac{x}{l_x} - \frac{\Omega t}{2}\right), \label{eqn:eps_mod}
\end{equation}
for positions such that $\mathbf{r}$ is in silicon. For simplicity we assume that the modulation occurs only for $|y| < l_y$, and $\Delta \varepsilon(\mathbf{r}, t) = 0$ otherwise. In what follows we set $l_y = 2\sqrt{3}a$, but we note that the results below do not depend qualitatively on this particular choice. In eq. (\ref{eqn:eps_mod}), we defined the real-space periodicity $l_x = 2\pi/\Delta_k$, as well as the maximum induced permittivity change $\Delta_\varepsilon$, relative to $\varepsilon_S$. 

In Fig. \ref{fig_schematic}(b), we show $\varepsilon(\mathbf{r}, t)$ in the plane at $z = 0$, at $t = 0$, $t = T/3$, and $t = 2T/3$, for $\Delta_\varepsilon = 0.3$ and $l_x = 16a$ (as before, $T = 2\pi/\Omega$). We note that the modulation amplitude is set to an unphysically high value in this panel only for illustrative purposes. This modulation thus imposes a supercell in the $x$-direction containing 16 unit cells of the unmodulated waveguide. This periodicity is preserved for all $t$, hence the Bloch momentum $k_x$ of light propagating along the modulated waveguide is still a conserved quantity, but the Brillouin zone is now folded to the region $k_x \in [-\Delta_k/2, \Delta_k/2]$. In Fig. \ref{fig_schematic}(d), we plot the instantaneous band structure at $t = 0$, computed with the guided-mode expansion with $L_x = l_x = 16a$ and $\Delta_\varepsilon = 0.03$. As can be seen, the lowest-frequency band is flattened by the modulated permittivity. Furthermore a band-gap separating it from all the higher-frequency bands is opened. This band structure is thus a promising starting point for implementing a unidirectional transport scheme analogous to that of Section \ref{sec:ccw}. Finally, we note that the modes of the lowest band are localized around the maximum of the permittivity distribution. This is illustrated in Fig. \ref{fig_schematic}(b), where we plot the electric field intensity of the lowest-frequency mode at $k_x = 0$ (the modes of this band look qualitatively similar at all $k_x$). The gradual modulation of the permittivity in essence creates a coupled-cavity waveguide, with a gently-confined photonic crystal cavity similar to the one of Ref. \cite{Song2005} at each node.

\subsection{Dynamic simulation}

The instantaneous eigenstates ($\mathbf{H}_{k\mu}(t)$, $\mathbf{E}_{k\mu}(t)$) at any time $t$ can be computed using the guided-mode expansion. We note that this method is approximate in that the coupling to modes in the radiative continuum  is only included perturbatively. However, this should be an extremely good approximation for the modes we study here, since they are formed by the part of the guided band of the underlying PhC that lies below the light cone (Fig. \ref{fig_schematic}(c)). Indeed, we obtain an extremely high quality factor $Q > 10^{12}$ for all modes of the lowest-frequency band shown in panel (d). Once the eigenmodes are computed over a discretized mesh in time, using the expansion in eqs. (\ref{eqn:u}) we can also simulate the full dynamics of a given starting state 
\begin{equation}
\bm{u}(\mathbf{r}, t = 0) = \sum_{k, \nu} e^{ikx} c_{k\nu}(0) \bm{u}_{k\nu}(\mathbf{r}, 0),
\end{equation}
defined by the expansion coefficients $c_{k\nu}(0)$. Everywhere below, we label the lowest guided band (e.g. the lowest brown band of Fig. \ref{fig_schematic}(d)) with the index $\nu = 1$, and consider a starting state that only contains modes in that band, i.e. $c_{k\nu}(0) \propto \delta_{\nu 1}$. We note that the dynamic simulation performed in this way is in principle exact, in the limit in which all (infinitely many) bands are included in the summation. As can be expected, for an adiabatic modulation, we find that the summation converges fast, with the strongest mixing occurring only within the few bands that are closest in frequency to the starting one.

\begin{figure}
\centering
\includegraphics[width = 0.48\textwidth, trim = 0in 0in 0in 0in, clip = true]{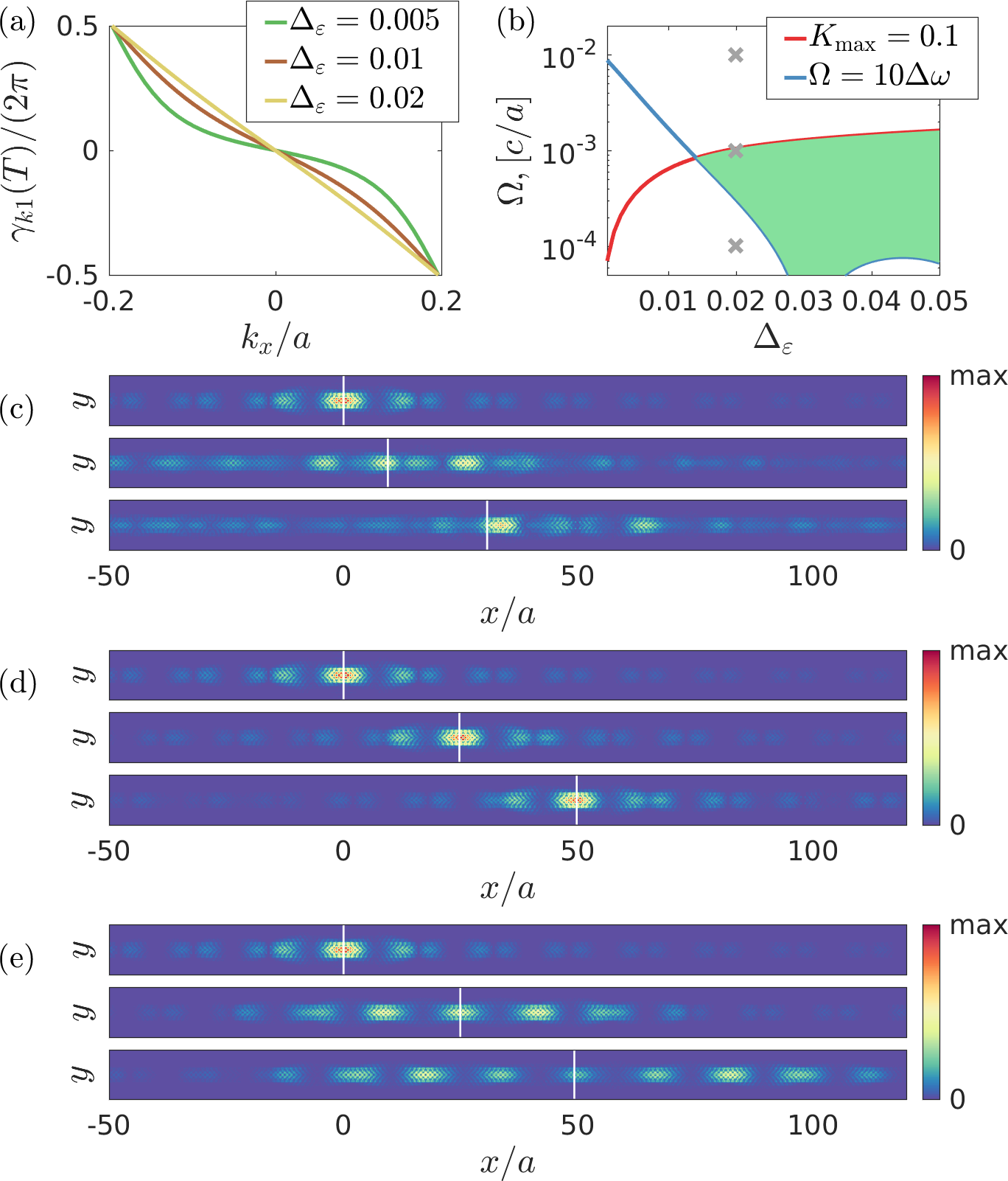}
 \caption{(a): Berry phase for the first guided band under a dynamic modulation with three different amplitudes. (b): The green shaded region shows the parameter space in modulation amplitude $\Delta_\varepsilon$ and frequency $\Omega$ within which unidirectional transport can be expected. Below the red line, the maximum overlap term $K_{\mathrm{max}}$ for the first band is much smaller than one. Above the blue line, the bandwidth $\Delta\omega_1$ is much smaller than $\Omega$. (c)-(e): Snapshots at times $t = 0$, $t/T = 25/16$, and $t/T = 50/16$ of the electric field $|E_x|^2 + |E_y|^2$ corresponding to the propagation of a wavepacket initially localized around $x = 0$, for the modulation parameters shown by crosses in panel (b), namely $\Delta_\varepsilon = 0.02$ and (c): $\Omega a/c = 10^{-2}$, (d): $\Omega a/c = 10^{-3}$, and (e): $\Omega a/c = 10^{-4}$, where $c$ denotes the speed of light. The vertical white lines show the position of the center of mass of the wavepacket at each time.} 
\label{fig_phc_wan}
\end{figure}

As we already discussed in Sections \ref{sec:floq_ad} and \ref{sec:ccw}, there are three requirements for unidirectional transport: winding Berry phase, adiabatic evolution, and a flat starting band on the scale of the modulation frequency $\Omega$. Thus, in Fig. \ref{fig_phc_wan}(a), we first plot the Berry phase $\gamma_{k1}(T)$ associated to the first band, computed as in eq. (\ref{eqn:phc_berry}), for $l_x = 16a$ and three different values of $\Delta_\varepsilon$. The sliding permittivity causes the Berry phase to wind in the Brillouin zone in the same way as in our system of Section \ref{sec:ccw}. Next, in Fig. \ref{fig_phc_wan}(b), we explore the parameter range in which the remaining two conditions are satisfied. We define, as before, $K_{\mathrm{max}} = \mathrm{max}_{k, t, \nu}(|K_{1\nu}(k, t)|)$, with $K_{1\nu}$ from eq. (\ref{eqn:phc_K}). The adiabatic condition is thus defined by $K_{\mathrm{max}} \ll 1$, or approximately within the region below the red line in Fig. \ref{fig_phc_wan}(a), which shows $K_{\mathrm{max}} = 0.1$. The blue line in the plot, on the other hand, delimitates the region in which the modulation frequency $\Omega$ is much larger than the bandwidth $\Delta \omega_1 = \mathrm{max}_k(\omega_{k1}) - \mathrm{min}_k(\omega_{k1})$. Thus, the green region between the two curves shows the target parameter space in $\Delta_\varepsilon$, $\Omega$, for the particular choice of $l_x = 16a$. 

In Fig. \ref{fig_phc_wan}(c)-(e), we plot the time evolution at three different times for a starting wavepacket centered around $x = 0$, assuming $\Delta_\varepsilon = 0.02$ everywhere, and $\Omega = 10^{-2} c/a$ in (c), $\Omega = 10^{-3} c/a$ in (d), and $\Omega = 10^{-4} c/a$ in (e) (parameters marked by crosses in panel (a)). In other words, the starting coefficients are given by 
\begin{equation}
 c_{k\nu}(0) = \delta_{1\nu}, \quad \forall k, 
 \end{equation} 
i.e. the first band is filled while the others are empty. In all panels, we also show with white vertical lines the position along the $x$-axis of the center of mass of the wavepacket, which we define as
\begin{equation}
x_\mathrm{com} = \frac{\int x\left(|E_x(x, 0)|^2 + |E_y(y, 0)|^2\right) \mathrm{d}x}{\int \left(|E_x(x, 0)|^2 + |E_y(y, 0)|^2\right) \mathrm{d}x},
\end{equation}
In panel (c), the evolution is not adiabatic -- the modulation frequency $\Omega$ is too high, and the mode is not well-guided. Namely, the wavepacket broadens, and its center of mass moves slower than the permittivity modulation. In contrast, both panels (d) and (e) represent adiabatic evolution, and are in fact an illustration of Thouless pumping. In both cases, the center of mass of the wavepacket slides together with the sliding potential (here, the permittivity). However, the difference between the two panels serves to once again illustrate the need for a sufficiently flat starting band (see panel (b)). In the case of panel (e), the modulation frequency $\Omega$ is too small compared to the bandwidth of the starting band $\omega_{k1}$. Thus, while the center of mass shifts adiabatically with the modulation, the mode broadens significantly. The best adiabatic guiding is thus observed in panel (d), in which the modulation parameters lie in the green region of panel (b). Using these parameters and starting wavepackets narrowly centered around a given $k_0$, we have further checked that the group velocity $\bar{v}(k_0)$ is constant throughout the whole Brillouin zone, confirming that the unidirectional light transport discussed in Sections \ref{sec:floq_ad} and \ref{sec:ccw} can also be achieved in the photonic crystal setup presented here. 

\begin{figure*}
\centering
\includegraphics[width = 0.8\textwidth, trim = 0in 0in 0in 0in, clip = true]{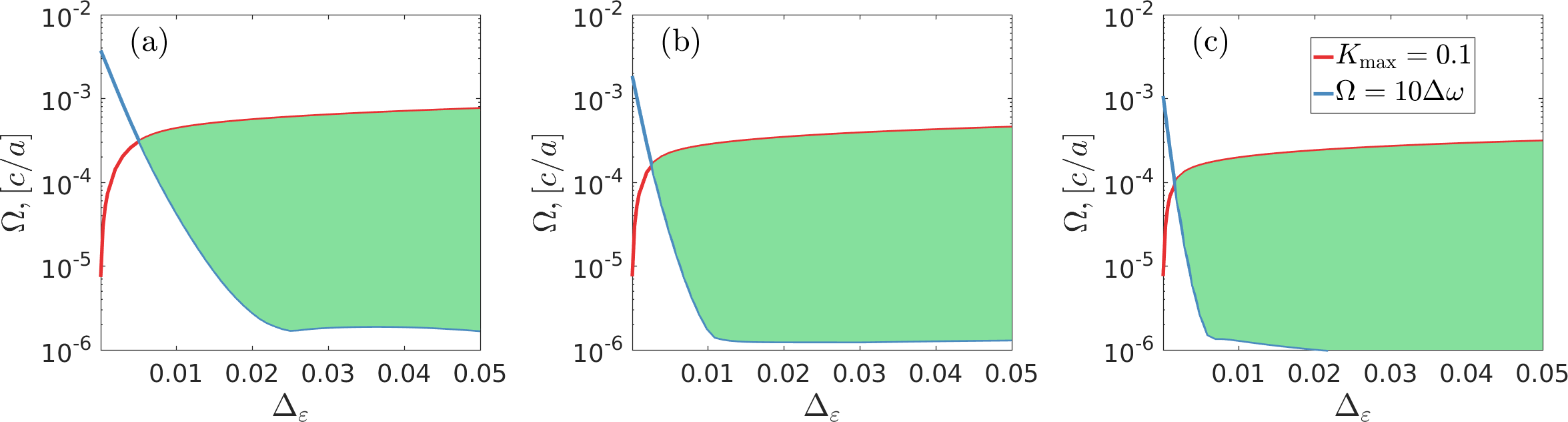}
 \caption{Same as Fig. \ref{fig_phc_wan}(b), but with (a): $l_x = 26a$; (b): $l_x = 36a$; and (c): $l_x = 46a$.} 
\label{fig_sweetspot}
\end{figure*}

\section{Discussion and conclusion}
\label{sec:conclusion}

\subsection{Experimental considerations}

So far, we used generic parameters expressed in units of the modulation frequency $\Omega$ in Section \ref{sec:ccw}, or $c/a$ in Section \ref{sec:phc}. Here, we discuss the experimental feasibility of the modulation parameters, as well as the group index of the slow light that can be expected in several sample structures. 

The group index for the adiabatic unidirectional guiding in both of our proposed implementations is given by 
\begin{equation}
n_g = c \frac{T}{L_c} = \frac{2\pi c}{\Omega L_c},
\label{eqn:ng}
\end{equation}
where $L_c$ is the distance travelled by a pulse within one cycle. The coupled-cavity waveguide of Section \ref{sec:ccw} is conceptually straightforward to implement with e.g. microring or microdisk cavities, which can be modulated at frequencies in the range of tens of GHz \cite{Xu2005, Janner2009, Reed2010}. For this system, $L_c = Nd_r$, where $N$ is the periodicity of the modulation defined in Section \ref{sec:ccw}, and $d_r$ is the center-to-center distance between nearest neighbor rings, which is approximately the ring diameter. Thus, as an example, for the modulation scheme used in Figs. \ref{fig_source}, \ref{fig_delay}, and \ref{fig_delay_dis}, assuming $d_r = 10\mu$m and a modulation frequency $\Omega/(2\pi) = 10$GHz, we compute through eq. (\ref{eqn:ng}) a group index $n_g = 188$. This is already in the range of the largest slow-light values ever reported \cite{Schulz2010}, and can be further increased by decreasing $\Omega$, or by decreasing the ring diameter. We note that a lower bound on the modulation frequency is set by the intrinsic loss $\kappa$ associated with each cavity, which has to be such that $\kappa \ll \Omega$. In our example, a quality factor of $Q = 10^{5}$ is sufficient, as it corresponds to a damping rate $\kappa = \omega/(2Q) \approx 1$GHz, assuming $\omega/(2\pi) \approx 200$THz. Thus, high-$Q$ cavities are needed, but the value is still two orders of magnitude smaller than what has been demonstrated in state-of the art silicon-based \cite{Tien2011, Asano2017} or lithium niobate \cite{Zhang2017} resonators.

For the PhC waveguide of Section \ref{sec:phc}, we simply have $L_c = l_x$. Using the parameters of Fig. \ref{fig_phc_wan}(d), namely $a = 400$nm, $l_x = 16a$, and $\Omega = 10^{-3}c/a$, i.e. $\Omega/(2\pi) \approx 119$GHz, we thus compute $n_g = 393$. We note, however, that this particular modulation frequency is challenging. Thus in Fig. \ref{fig_sweetspot} we further explore the parameter range in order to find experimentally accessible parameters within which adiabatic guiding is possible. We also note that the particular example of a modulation scheme that we studied in Section \ref{sec:phc} was only taken for concreteness, but other schemes could also apply, including electro-optic modulation as in \cite{Nguyen2011, Lira2012}, or using the $\chi^{(2)}$-nonlinearity of materials like lithium niobate \cite{Janner2009, Wang2017}.  

In Fig. \ref{fig_sweetspot}, we show how the range in parameter space with which adiabatic guiding can be achieved depends on the modulation periodicity $l_x$. Generally speaking, we observe that the green region shifts towards lower $\Omega$ and $\Delta_\varepsilon$ with increasing $l_x$. Thus, for example for $l_x = 36$ (Fig. \ref{fig_sweetspot}(b)), adiabatic guiding can be achieved for $\Omega/(2\pi) = 10$GHz, i.e. $\Omega/(2\pi) = 1.33 \times 10^{-5}(c/a)$, and for $\Delta_\varepsilon = 0.004$, i.e a relative refractive index change in the material of $\Delta n/n = 0.2\%$. These parameters are reasonable for state-of-the-art technologies, and the group index computed through eq. (\ref{eqn:ng}) is $n_g = 2,089$. For the waveguide with $l_x = 46a$ of Fig. \ref{fig_sweetspot}(c), and the same value of $\Delta_\varepsilon = 0.004$, we can have $\Omega/(2\pi) = 1$GHz, corresponding to a group index of $n_g = 16,345$. Such high group indexes are possible in the PhC implementation because of the extremely compact field concentration, i.e. very short $l_x$. Our scheme thus provides a way to radically overcome the group-index limit set by back-scattering in standard photonic structures. 

\subsection{Protection against disorder}

It is important to highlight that the protection against disorder shown in Fig. \ref{fig_delay_dis} is of a completely novel nature. In particular, we have managed to decouple the group index, $n_{g} \propto 1/(\Omega L_c)$, from the maximum disorder magnitude $\sigma$ for which transport persists. In other words, for any arbitrary group velocity that can be set by controlling $\Omega$ and/or $L_c$, we can in principle have arbitrarily large disorder protection by increasing $J$ and $A_0$. This is in sharp contrast with the case of a regular slow-light CROW (as in Fig. \ref{fig_delay}(a) and \ref{fig_delay}(b)). In such a device, setting the group velocity \textit{directly} sets a limitation on the maximum allowed disorder. This is because $v_g \propto J_s$, where $J_s$ is the slow-light coupling constant, and $\sigma \ll J_s$ is required for operation. 

Our system should also be contrasted with the case of photonic topological insulators achieved through dynamic modulation \cite{Fang2012, Minkov2016}. In these systems, one-way Floquet bands have also been predicted, but the topological band gap is inevitably given by a fraction of the modulation frequency $\Omega$. The size of this band gap is in fact what determines the magnitude of the disorder protection, which means that in these systems $\sigma \ll \Omega$ is required. On the other hand, the band gap also determines the bandwidth of the one-way edge state, and hence the group velocity is again proportional to $\Omega$, as well as to the spatial periodicity. In short, however, just as in the case of the standard CROW, the maximum disorder and the maximum group index are again related, only this time through $\Omega$ instead of through $J_s$. Furthermore, in certain systems, the requirement $\sigma \ll \Omega$  could be much harder to achieve compared to $\sigma \ll J, A_0$, which is needed in our system. Thus, we have uncovered a regime of protection against disorder which is fundamentally different from the effect associated to Floquet topological insulators, and which leads in certain cases to \textit{stronger} protection, despite the fact that it is achieved in a purely one-dimensional system. 

\subsection{Conclusion}

In conclusion, we have proposed and extensively studied a paradigm for unidirectional light transport in a one-dimensional waveguide that can be achieved through dynamic modulation. The theoretical considerations in Section \ref{sec:floq_ad} provide some general insights that apply to various systems, including outside the domain of optics, like cold-atom \cite{Jotzu2014} or acoustic \cite{Khanikaev2015} platforms. The particular examples given in Sections \ref{sec:ccw} and \ref{sec:phc} use a modulated CROW and a modulated PhC waveguide, respectively, both of which are standard building blocks of integrated photonic devices. We have identified a range of possible parameters that achieve the unidirectional transport, and we have shown that a part of this range falls within what can be implemented in state-of-the-art photonic technologies. On the fundamental level, we have identified a conceptually novel regime of disorder protection. In particular, we have demonstrated that the robustness with respect to imperfections that is a hallmark of two-dimensional photonic topological insulators can also be achieved in one-dimensional dynamically modulated systems. This could significantly strengthen the significance of the robust transport for practical applications.

This work was supported by the Swiss National Science Foundation through Project N\textsuperscript{\underline{o}} P300P2\_177721, and the US Air Force Office of Scientific Research FA9550-17-1-0002.


\begin{thebibliography}{44}%
\makeatletter
\providecommand \@ifxundefined [1]{%
 \@ifx{#1\undefined}
}%
\providecommand \@ifnum [1]{%
 \ifnum #1\expandafter \@firstoftwo
 \else \expandafter \@secondoftwo
 \fi
}%
\providecommand \@ifx [1]{%
 \ifx #1\expandafter \@firstoftwo
 \else \expandafter \@secondoftwo
 \fi
}%
\providecommand \natexlab [1]{#1}%
\providecommand \enquote  [1]{``#1''}%
\providecommand \bibnamefont  [1]{#1}%
\providecommand \bibfnamefont [1]{#1}%
\providecommand \citenamefont [1]{#1}%
\providecommand \href@noop [0]{\@secondoftwo}%
\providecommand \href [0]{\begingroup \@sanitize@url \@href}%
\providecommand \@href[1]{\@@startlink{#1}\@@href}%
\providecommand \@@href[1]{\endgroup#1\@@endlink}%
\providecommand \@sanitize@url [0]{\catcode `\\12\catcode `\$12\catcode
  `\&12\catcode `\#12\catcode `\^12\catcode `\_12\catcode `\%12\relax}%
\providecommand \@@startlink[1]{}%
\providecommand \@@endlink[0]{}%
\providecommand \url  [0]{\begingroup\@sanitize@url \@url }%
\providecommand \@url [1]{\endgroup\@href {#1}{\urlprefix }}%
\providecommand \urlprefix  [0]{URL }%
\providecommand \Eprint [0]{\href }%
\providecommand \doibase [0]{http://dx.doi.org/}%
\providecommand \selectlanguage [0]{\@gobble}%
\providecommand \bibinfo  [0]{\@secondoftwo}%
\providecommand \bibfield  [0]{\@secondoftwo}%
\providecommand \translation [1]{[#1]}%
\providecommand \BibitemOpen [0]{}%
\providecommand \bibitemStop [0]{}%
\providecommand \bibitemNoStop [0]{.\EOS\space}%
\providecommand \EOS [0]{\spacefactor3000\relax}%
\providecommand \BibitemShut  [1]{\csname bibitem#1\endcsname}%
\let\auto@bib@innerbib\@empty
\bibitem [{\citenamefont {Krauss}(2008)}]{Krauss2008}%
  \BibitemOpen
  \bibfield  {author} {\bibinfo {author} {\bibfnamefont {T.~F.}\ \bibnamefont
  {Krauss}},\ }\href@noop {} {\bibfield  {journal} {\bibinfo  {journal} {Nat.
  Photonics}\ }\textbf {\bibinfo {volume} {2}},\ \bibinfo {pages} {448}
  (\bibinfo {year} {2008})}\BibitemShut {NoStop}%
\bibitem [{\citenamefont {Baba}(2008)}]{Baba2008}%
  \BibitemOpen
  \bibfield  {author} {\bibinfo {author} {\bibfnamefont {T.}~\bibnamefont
  {Baba}},\ }\href@noop {} {\bibfield  {journal} {\bibinfo  {journal} {Nat.
  Photonics}\ }\textbf {\bibinfo {volume} {2}},\ \bibinfo {pages} {465}
  (\bibinfo {year} {2008})}\BibitemShut {NoStop}%
\bibitem [{\citenamefont {Schulz}\ \emph {et~al.}(2010)\citenamefont {Schulz},
  \citenamefont {O'Faolain}, \citenamefont {Beggs}, \citenamefont {White},
  \citenamefont {Melloni},\ and\ \citenamefont {Krauss}}]{Schulz2010}%
  \BibitemOpen
  \bibfield  {author} {\bibinfo {author} {\bibfnamefont {S.~A.}\ \bibnamefont
  {Schulz}}, \bibinfo {author} {\bibfnamefont {L.}~\bibnamefont {O'Faolain}},
  \bibinfo {author} {\bibfnamefont {D.~M.}\ \bibnamefont {Beggs}}, \bibinfo
  {author} {\bibfnamefont {T.~P.}\ \bibnamefont {White}}, \bibinfo {author}
  {\bibfnamefont {A.}~\bibnamefont {Melloni}}, \ and\ \bibinfo {author}
  {\bibfnamefont {T.~F.}\ \bibnamefont {Krauss}},\ }\href@noop {} {\bibfield
  {journal} {\bibinfo  {journal} {Journal of Optics}\ }\textbf {\bibinfo
  {volume} {12}},\ \bibinfo {pages} {104004} (\bibinfo {year}
  {2010})}\BibitemShut {NoStop}%
\bibitem [{\citenamefont {Yariv}\ \emph {et~al.}(1999)\citenamefont {Yariv},
  \citenamefont {Xu}, \citenamefont {Lee},\ and\ \citenamefont
  {Scherer}}]{Yariv1999}%
  \BibitemOpen
  \bibfield  {author} {\bibinfo {author} {\bibfnamefont {A.}~\bibnamefont
  {Yariv}}, \bibinfo {author} {\bibfnamefont {Y.}~\bibnamefont {Xu}}, \bibinfo
  {author} {\bibfnamefont {R.~K.}\ \bibnamefont {Lee}}, \ and\ \bibinfo
  {author} {\bibfnamefont {A.}~\bibnamefont {Scherer}},\ }\href@noop {}
  {\bibfield  {journal} {\bibinfo  {journal} {Opt. Lett.}\ }\textbf {\bibinfo
  {volume} {24}},\ \bibinfo {pages} {711} (\bibinfo {year} {1999})}\BibitemShut
  {NoStop}%
\bibitem [{\citenamefont {Patterson}\ \emph {et~al.}(2009)\citenamefont
  {Patterson}, \citenamefont {Hughes}, \citenamefont {Combri{\'{e}}},
  \citenamefont {Tran}, \citenamefont {{De Rossi}}, \citenamefont {Gabet},\
  and\ \citenamefont {Jaou{\"{e}}n}}]{Patterson2009}%
  \BibitemOpen
  \bibfield  {author} {\bibinfo {author} {\bibfnamefont {M.}~\bibnamefont
  {Patterson}}, \bibinfo {author} {\bibfnamefont {S.}~\bibnamefont {Hughes}},
  \bibinfo {author} {\bibfnamefont {S.}~\bibnamefont {Combri{\'{e}}}}, \bibinfo
  {author} {\bibfnamefont {N.~V.~Q.}\ \bibnamefont {Tran}}, \bibinfo {author}
  {\bibfnamefont {A.}~\bibnamefont {{De Rossi}}}, \bibinfo {author}
  {\bibfnamefont {R.}~\bibnamefont {Gabet}}, \ and\ \bibinfo {author}
  {\bibfnamefont {Y.}~\bibnamefont {Jaou{\"{e}}n}},\ }\href@noop {} {\bibfield
  {journal} {\bibinfo  {journal} {Phys. Rev. Lett.}\ }\textbf {\bibinfo
  {volume} {102}},\ \bibinfo {pages} {1} (\bibinfo {year} {2009})}\BibitemShut
  {NoStop}%
\bibitem [{\citenamefont {Mazoyer}\ \emph {et~al.}(2009)\citenamefont
  {Mazoyer}, \citenamefont {Hugonin},\ and\ \citenamefont
  {Lalanne}}]{Mazoyer2009}%
  \BibitemOpen
  \bibfield  {author} {\bibinfo {author} {\bibfnamefont {S.}~\bibnamefont
  {Mazoyer}}, \bibinfo {author} {\bibfnamefont {J.}~\bibnamefont {Hugonin}}, \
  and\ \bibinfo {author} {\bibfnamefont {P.}~\bibnamefont {Lalanne}},\
  }\href@noop {} {\bibfield  {journal} {\bibinfo  {journal} {Phys. Rev. Lett.}\
  }\textbf {\bibinfo {volume} {103}},\ \bibinfo {pages} {063903} (\bibinfo
  {year} {2009})}\BibitemShut {NoStop}%
\bibitem [{\citenamefont {Minkov}\ and\ \citenamefont
  {Savona}(2013)}]{Minkov2013}%
  \BibitemOpen
  \bibfield  {author} {\bibinfo {author} {\bibfnamefont {M.}~\bibnamefont
  {Minkov}}\ and\ \bibinfo {author} {\bibfnamefont {V.}~\bibnamefont
  {Savona}},\ }\href@noop {} {\bibfield  {journal} {\bibinfo  {journal} {Phys.
  Rev. B}\ }\textbf {\bibinfo {volume} {88}},\ \bibinfo {pages} {081303}
  (\bibinfo {year} {2013})}\BibitemShut {NoStop}%
\bibitem [{\citenamefont {John}(1987)}]{John1987}%
  \BibitemOpen
  \bibfield  {author} {\bibinfo {author} {\bibfnamefont {S.}~\bibnamefont
  {John}},\ }\href@noop {} {\bibfield  {journal} {\bibinfo  {journal} {Phys.
  Rev. Lett.}\ }\textbf {\bibinfo {volume} {58}},\ \bibinfo {pages} {2486}
  (\bibinfo {year} {1987})}\BibitemShut {NoStop}%
\bibitem [{\citenamefont {Lu}\ \emph {et~al.}(2014)\citenamefont {Lu},
  \citenamefont {Joannopoulos},\ and\ \citenamefont
  {Solja\v{c}i\'{c}}}]{Lu2014}%
  \BibitemOpen
  \bibfield  {author} {\bibinfo {author} {\bibfnamefont {L.}~\bibnamefont
  {Lu}}, \bibinfo {author} {\bibfnamefont {J.~D.}\ \bibnamefont
  {Joannopoulos}}, \ and\ \bibinfo {author} {\bibfnamefont {M.}~\bibnamefont
  {Solja\v{c}i\'{c}}},\ }\href@noop {} {\bibfield  {journal} {\bibinfo
  {journal} {Nat. Photonics}\ }\textbf {\bibinfo {volume} {8}},\ \bibinfo
  {pages} {821} (\bibinfo {year} {2014})}\BibitemShut {NoStop}%
\bibitem [{\citenamefont {Raghu}\ and\ \citenamefont
  {Haldane}(2008)}]{Raghu2008}%
  \BibitemOpen
  \bibfield  {author} {\bibinfo {author} {\bibfnamefont {S.}~\bibnamefont
  {Raghu}}\ and\ \bibinfo {author} {\bibfnamefont {F.}~\bibnamefont
  {Haldane}},\ }\href@noop {} {\bibfield  {journal} {\bibinfo  {journal} {Phys.
  Rev. A}\ }\textbf {\bibinfo {volume} {78}},\ \bibinfo {pages} {033834}
  (\bibinfo {year} {2008})}\BibitemShut {NoStop}%
\bibitem [{\citenamefont {Haldane}\ and\ \citenamefont
  {Raghu}(2008)}]{Haldane2008}%
  \BibitemOpen
  \bibfield  {author} {\bibinfo {author} {\bibfnamefont {F.}~\bibnamefont
  {Haldane}}\ and\ \bibinfo {author} {\bibfnamefont {S.}~\bibnamefont
  {Raghu}},\ }\href@noop {} {\bibfield  {journal} {\bibinfo  {journal} {Phys.
  Rev. Lett.}\ }\textbf {\bibinfo {volume} {100}},\ \bibinfo {pages} {013904}
  (\bibinfo {year} {2008})}\BibitemShut {NoStop}%
\bibitem [{\citenamefont {Wang}\ \emph {et~al.}(2009)\citenamefont {Wang},
  \citenamefont {Chong}, \citenamefont {Joannopoulos},\ and\ \citenamefont
  {Soljaci\'{c}}}]{Wang2009}%
  \BibitemOpen
  \bibfield  {author} {\bibinfo {author} {\bibfnamefont {Z.}~\bibnamefont
  {Wang}}, \bibinfo {author} {\bibfnamefont {Y.}~\bibnamefont {Chong}},
  \bibinfo {author} {\bibfnamefont {J.~D.}\ \bibnamefont {Joannopoulos}}, \
  and\ \bibinfo {author} {\bibfnamefont {M.}~\bibnamefont {Soljaci\'{c}}},\
  }\href@noop {} {\bibfield  {journal} {\bibinfo  {journal} {Nature}\ }\textbf
  {\bibinfo {volume} {461}},\ \bibinfo {pages} {772} (\bibinfo {year}
  {2009})}\BibitemShut {NoStop}%
\bibitem [{\citenamefont {Hafezi}\ \emph {et~al.}(2011)\citenamefont {Hafezi},
  \citenamefont {Demler}, \citenamefont {Lukin},\ and\ \citenamefont
  {Taylor}}]{Hafezi2011}%
  \BibitemOpen
  \bibfield  {author} {\bibinfo {author} {\bibfnamefont {M.}~\bibnamefont
  {Hafezi}}, \bibinfo {author} {\bibfnamefont {E.~a.}\ \bibnamefont {Demler}},
  \bibinfo {author} {\bibfnamefont {M.~D.}\ \bibnamefont {Lukin}}, \ and\
  \bibinfo {author} {\bibfnamefont {J.~M.}\ \bibnamefont {Taylor}},\
  }\href@noop {} {\bibfield  {journal} {\bibinfo  {journal} {Nat. Phys.}\
  }\textbf {\bibinfo {volume} {7}},\ \bibinfo {pages} {907} (\bibinfo {year}
  {2011})}\BibitemShut {NoStop}%
\bibitem [{\citenamefont {Rechtsman}\ \emph {et~al.}(2013)\citenamefont
  {Rechtsman}, \citenamefont {Zeuner}, \citenamefont {Plotnik}, \citenamefont
  {Lumer}, \citenamefont {Podolsky}, \citenamefont {Dreisow}, \citenamefont
  {Nolte}, \citenamefont {Segev},\ and\ \citenamefont
  {Szameit}}]{Rechtsman2013}%
  \BibitemOpen
  \bibfield  {author} {\bibinfo {author} {\bibfnamefont {M.~C.}\ \bibnamefont
  {Rechtsman}}, \bibinfo {author} {\bibfnamefont {J.~M.}\ \bibnamefont
  {Zeuner}}, \bibinfo {author} {\bibfnamefont {Y.}~\bibnamefont {Plotnik}},
  \bibinfo {author} {\bibfnamefont {Y.}~\bibnamefont {Lumer}}, \bibinfo
  {author} {\bibfnamefont {D.}~\bibnamefont {Podolsky}}, \bibinfo {author}
  {\bibfnamefont {F.}~\bibnamefont {Dreisow}}, \bibinfo {author} {\bibfnamefont
  {S.}~\bibnamefont {Nolte}}, \bibinfo {author} {\bibfnamefont
  {M.}~\bibnamefont {Segev}}, \ and\ \bibinfo {author} {\bibfnamefont
  {A.}~\bibnamefont {Szameit}},\ }\href@noop {} {\bibfield  {journal} {\bibinfo
   {journal} {Nature}\ }\textbf {\bibinfo {volume} {496}},\ \bibinfo {pages}
  {196} (\bibinfo {year} {2013})}\BibitemShut {NoStop}%
\bibitem [{\citenamefont {Fang}\ \emph {et~al.}(2012)\citenamefont {Fang},
  \citenamefont {Yu},\ and\ \citenamefont {Fan}}]{Fang2012}%
  \BibitemOpen
  \bibfield  {author} {\bibinfo {author} {\bibfnamefont {K.}~\bibnamefont
  {Fang}}, \bibinfo {author} {\bibfnamefont {Z.}~\bibnamefont {Yu}}, \ and\
  \bibinfo {author} {\bibfnamefont {S.}~\bibnamefont {Fan}},\ }\href@noop {}
  {\bibfield  {journal} {\bibinfo  {journal} {Nat. Photonics}\ }\textbf
  {\bibinfo {volume} {6}},\ \bibinfo {pages} {782} (\bibinfo {year}
  {2012})}\BibitemShut {NoStop}%
\bibitem [{\citenamefont {Minkov}\ and\ \citenamefont
  {Savona}(2016)}]{Minkov2016}%
  \BibitemOpen
  \bibfield  {author} {\bibinfo {author} {\bibfnamefont {M.}~\bibnamefont
  {Minkov}}\ and\ \bibinfo {author} {\bibfnamefont {V.}~\bibnamefont
  {Savona}},\ }\href@noop {} {\bibfield  {journal} {\bibinfo  {journal}
  {Optica}\ }\textbf {\bibinfo {volume} {3}},\ \bibinfo {pages} {200} (\bibinfo
  {year} {2016})}\BibitemShut {NoStop}%
\bibitem [{\citenamefont {Bahari}\ \emph {et~al.}(2017)\citenamefont {Bahari},
  \citenamefont {Ndao}, \citenamefont {Vallini}, \citenamefont {El~Amili},
  \citenamefont {Fainman},\ and\ \citenamefont {Kant{\'e}}}]{Bahari2017}%
  \BibitemOpen
  \bibfield  {author} {\bibinfo {author} {\bibfnamefont {B.}~\bibnamefont
  {Bahari}}, \bibinfo {author} {\bibfnamefont {A.}~\bibnamefont {Ndao}},
  \bibinfo {author} {\bibfnamefont {F.}~\bibnamefont {Vallini}}, \bibinfo
  {author} {\bibfnamefont {A.}~\bibnamefont {El~Amili}}, \bibinfo {author}
  {\bibfnamefont {Y.}~\bibnamefont {Fainman}}, \ and\ \bibinfo {author}
  {\bibfnamefont {B.}~\bibnamefont {Kant{\'e}}},\ }\href@noop {} {\bibfield
  {journal} {\bibinfo  {journal} {Science}\ }\textbf {\bibinfo {volume}
  {358}},\ \bibinfo {pages} {636–640} (\bibinfo {year} {2017})}\BibitemShut
  {NoStop}%
\bibitem [{\citenamefont {Shirley}(1965)}]{Shirley1965}%
  \BibitemOpen
  \bibfield  {author} {\bibinfo {author} {\bibfnamefont {J.~H.}\ \bibnamefont
  {Shirley}},\ }\href@noop {} {\bibfield  {journal} {\bibinfo  {journal} {Phys.
  Rev.}\ }\textbf {\bibinfo {volume} {138}},\ \bibinfo {pages} {B979} (\bibinfo
  {year} {1965})}\BibitemShut {NoStop}%
\bibitem [{\citenamefont {Kitagawa}\ \emph {et~al.}(2010)\citenamefont
  {Kitagawa}, \citenamefont {Berg}, \citenamefont {Rudner},\ and\ \citenamefont
  {Demler}}]{Kitagawa2010}%
  \BibitemOpen
  \bibfield  {author} {\bibinfo {author} {\bibfnamefont {T.}~\bibnamefont
  {Kitagawa}}, \bibinfo {author} {\bibfnamefont {E.}~\bibnamefont {Berg}},
  \bibinfo {author} {\bibfnamefont {M.}~\bibnamefont {Rudner}}, \ and\ \bibinfo
  {author} {\bibfnamefont {E.}~\bibnamefont {Demler}},\ }\href@noop {}
  {\bibfield  {journal} {\bibinfo  {journal} {Phys. Rev. B}\ }\textbf {\bibinfo
  {volume} {82}},\ \bibinfo {pages} {235114} (\bibinfo {year}
  {2010})}\BibitemShut {NoStop}%
\bibitem [{\citenamefont {Rudner}\ \emph {et~al.}(2013)\citenamefont {Rudner},
  \citenamefont {Lindner}, \citenamefont {Berg},\ and\ \citenamefont
  {Levin}}]{Rudner2013}%
  \BibitemOpen
  \bibfield  {author} {\bibinfo {author} {\bibfnamefont {M.~S.}\ \bibnamefont
  {Rudner}}, \bibinfo {author} {\bibfnamefont {N.~H.}\ \bibnamefont {Lindner}},
  \bibinfo {author} {\bibfnamefont {E.}~\bibnamefont {Berg}}, \ and\ \bibinfo
  {author} {\bibfnamefont {M.}~\bibnamefont {Levin}},\ }\href@noop {}
  {\bibfield  {journal} {\bibinfo  {journal} {Physical Review X}\ }\textbf
  {\bibinfo {volume} {3}},\ \bibinfo {pages} {031005} (\bibinfo {year}
  {2013})}\BibitemShut {NoStop}%
\bibitem [{\citenamefont {Thouless}(1983)}]{Thouless1983}%
  \BibitemOpen
  \bibfield  {author} {\bibinfo {author} {\bibfnamefont {D.~J.}\ \bibnamefont
  {Thouless}},\ }\href@noop {} {\bibfield  {journal} {\bibinfo  {journal}
  {Phys. Rev. B}\ }\textbf {\bibinfo {volume} {27}},\ \bibinfo {pages} {6083}
  (\bibinfo {year} {1983})}\BibitemShut {NoStop}%
\bibitem [{\citenamefont {Sakurai}(1994)}]{Sakurai1994}%
  \BibitemOpen
  \bibfield  {author} {\bibinfo {author} {\bibfnamefont {J.~J.}\ \bibnamefont
  {Sakurai}},\ }\href@noop {} {\emph {\bibinfo {title} {Modern Quantum
  Mechanics Revised Edition}}},\ edited by\ \bibinfo {editor} {\bibfnamefont
  {S.~F.}\ \bibnamefont {Tuan}}\ (\bibinfo  {publisher} {Adison-Wesley},\
  \bibinfo {year} {1994})\ Chap.~\bibinfo {chapter} {5}\BibitemShut {NoStop}%
\bibitem [{\citenamefont {Berry}(1984)}]{Berry1984}%
  \BibitemOpen
  \bibfield  {author} {\bibinfo {author} {\bibfnamefont {M.~V.}\ \bibnamefont
  {Berry}},\ }\href@noop {} {\bibfield  {journal} {\bibinfo  {journal}
  {Proceedings of the Royal Society of London A}\ }\textbf {\bibinfo {volume}
  {392}},\ \bibinfo {pages} {45} (\bibinfo {year} {1984})}\BibitemShut
  {NoStop}%
\bibitem [{\citenamefont {Xiao}\ \emph {et~al.}(2010)\citenamefont {Xiao},
  \citenamefont {Chang},\ and\ \citenamefont {Niu}}]{Xiao2010}%
  \BibitemOpen
  \bibfield  {author} {\bibinfo {author} {\bibfnamefont {D.}~\bibnamefont
  {Xiao}}, \bibinfo {author} {\bibfnamefont {M.-C.}\ \bibnamefont {Chang}}, \
  and\ \bibinfo {author} {\bibfnamefont {Q.}~\bibnamefont {Niu}},\ }\href@noop
  {} {\bibfield  {journal} {\bibinfo  {journal} {Reviews of Modern Physics}\
  }\textbf {\bibinfo {volume} {82}},\ \bibinfo {pages} {1959} (\bibinfo {year}
  {2010})}\BibitemShut {NoStop}%
\bibitem [{\citenamefont {Haus}(1984)}]{Haus1984}%
  \BibitemOpen
  \bibfield  {author} {\bibinfo {author} {\bibfnamefont {H.}~\bibnamefont
  {Haus}},\ }\href@noop {} {\emph {\bibinfo {title} {Waves and fields in
  optoelectronics}}},\ Prentice-Hall Series in Solid State Physical
  Electronics\ (\bibinfo  {publisher} {Prentice Hall, Incorporated},\ \bibinfo
  {year} {1984})\BibitemShut {NoStop}%
\bibitem [{\citenamefont {Fan}\ \emph {et~al.}(2003)\citenamefont {Fan},
  \citenamefont {Suh},\ and\ \citenamefont {Joannopoulos}}]{Fan2003}%
  \BibitemOpen
  \bibfield  {author} {\bibinfo {author} {\bibfnamefont {S.}~\bibnamefont
  {Fan}}, \bibinfo {author} {\bibfnamefont {W.}~\bibnamefont {Suh}}, \ and\
  \bibinfo {author} {\bibfnamefont {J.~D.}\ \bibnamefont {Joannopoulos}},\
  }\href@noop {} {\bibfield  {journal} {\bibinfo  {journal} {J. Opt. Soc. Am.
  A}\ }\textbf {\bibinfo {volume} {20}},\ \bibinfo {pages} {569} (\bibinfo
  {year} {2003})}\BibitemShut {NoStop}%
\bibitem [{\citenamefont {Minkov}\ \emph {et~al.}(2017)\citenamefont {Minkov},
  \citenamefont {Shi},\ and\ \citenamefont {Fan}}]{Minkov2017}%
  \BibitemOpen
  \bibfield  {author} {\bibinfo {author} {\bibfnamefont {M.}~\bibnamefont
  {Minkov}}, \bibinfo {author} {\bibfnamefont {Y.}~\bibnamefont {Shi}}, \ and\
  \bibinfo {author} {\bibfnamefont {S.}~\bibnamefont {Fan}},\ }\href@noop {}
  {\bibfield  {journal} {\bibinfo  {journal} {APL Photonics}\ }\textbf
  {\bibinfo {volume} {2}},\ \bibinfo {pages} {076101} (\bibinfo {year}
  {2017})}\BibitemShut {NoStop}%
\bibitem [{\citenamefont {Sumetsky}\ and\ \citenamefont
  {Eggleton}(2003)}]{Sumetsky2003}%
  \BibitemOpen
  \bibfield  {author} {\bibinfo {author} {\bibfnamefont {M.}~\bibnamefont
  {Sumetsky}}\ and\ \bibinfo {author} {\bibfnamefont {B.}~\bibnamefont
  {Eggleton}},\ }\href@noop {} {\bibfield  {journal} {\bibinfo  {journal} {Opt.
  Express}\ }\textbf {\bibinfo {volume} {11}},\ \bibinfo {pages} {381}
  (\bibinfo {year} {2003})}\BibitemShut {NoStop}%
\bibitem [{\citenamefont {Andreani}\ and\ \citenamefont
  {Gerace}(2006)}]{Andreani2006}%
  \BibitemOpen
  \bibfield  {author} {\bibinfo {author} {\bibfnamefont {L.~C.}\ \bibnamefont
  {Andreani}}\ and\ \bibinfo {author} {\bibfnamefont {D.}~\bibnamefont
  {Gerace}},\ }\href@noop {} {\bibfield  {journal} {\bibinfo  {journal} {Phys.
  Rev. B}\ }\textbf {\bibinfo {volume} {73}},\ \bibinfo {pages} {235114}
  (\bibinfo {year} {2006})}\BibitemShut {NoStop}%
\bibitem [{\citenamefont {Minkov}\ and\ \citenamefont
  {Savona}(2015)}]{Minkov2015}%
  \BibitemOpen
  \bibfield  {author} {\bibinfo {author} {\bibfnamefont {M.}~\bibnamefont
  {Minkov}}\ and\ \bibinfo {author} {\bibfnamefont {V.}~\bibnamefont
  {Savona}},\ }\href@noop {} {\bibfield  {journal} {\bibinfo  {journal}
  {Optica}\ }\textbf {\bibinfo {volume} {2}},\ \bibinfo {pages} {631} (\bibinfo
  {year} {2015})}\BibitemShut {NoStop}%
\bibitem [{\citenamefont {Leonard}\ \emph {et~al.}(2002)\citenamefont
  {Leonard}, \citenamefont {van Driel}, \citenamefont {Schilling},\ and\
  \citenamefont {Wehrspohn}}]{Leonard2002}%
  \BibitemOpen
  \bibfield  {author} {\bibinfo {author} {\bibfnamefont {S.~W.}\ \bibnamefont
  {Leonard}}, \bibinfo {author} {\bibfnamefont {H.~M.}\ \bibnamefont {van
  Driel}}, \bibinfo {author} {\bibfnamefont {J.}~\bibnamefont {Schilling}}, \
  and\ \bibinfo {author} {\bibfnamefont {R.~B.}\ \bibnamefont {Wehrspohn}},\
  }\href@noop {} {\bibfield  {journal} {\bibinfo  {journal} {Phys. Rev. B}\
  }\textbf {\bibinfo {volume} {66}},\ \bibinfo {pages} {161102} (\bibinfo
  {year} {2002})}\BibitemShut {NoStop}%
\bibitem [{\citenamefont {Beggs}\ \emph {et~al.}(2012)\citenamefont {Beggs},
  \citenamefont {Krauss}, \citenamefont {Kuipers},\ and\ \citenamefont
  {Kampfrath}}]{Beggs2012}%
  \BibitemOpen
  \bibfield  {author} {\bibinfo {author} {\bibfnamefont {D.~M.}\ \bibnamefont
  {Beggs}}, \bibinfo {author} {\bibfnamefont {T.~F.}\ \bibnamefont {Krauss}},
  \bibinfo {author} {\bibfnamefont {L.}~\bibnamefont {Kuipers}}, \ and\
  \bibinfo {author} {\bibfnamefont {T.}~\bibnamefont {Kampfrath}},\ }\href@noop
  {} {\bibfield  {journal} {\bibinfo  {journal} {Phys. Rev. Lett.}\ }\textbf
  {\bibinfo {volume} {108}},\ \bibinfo {pages} {033902} (\bibinfo {year}
  {2012})}\BibitemShut {NoStop}%
\bibitem [{\citenamefont {Song}\ \emph {et~al.}(2005)\citenamefont {Song},
  \citenamefont {Noda}, \citenamefont {Asano},\ and\ \citenamefont
  {Akahane}}]{Song2005}%
  \BibitemOpen
  \bibfield  {author} {\bibinfo {author} {\bibfnamefont {B.-S.}\ \bibnamefont
  {Song}}, \bibinfo {author} {\bibfnamefont {S.}~\bibnamefont {Noda}}, \bibinfo
  {author} {\bibfnamefont {T.}~\bibnamefont {Asano}}, \ and\ \bibinfo {author}
  {\bibfnamefont {Y.}~\bibnamefont {Akahane}},\ }\href@noop {} {\bibfield
  {journal} {\bibinfo  {journal} {Nat. Mater.}\ }\textbf {\bibinfo {volume}
  {4}},\ \bibinfo {pages} {207} (\bibinfo {year} {2005})}\BibitemShut {NoStop}%
\bibitem [{\citenamefont {Xu}\ \emph {et~al.}(2005)\citenamefont {Xu},
  \citenamefont {Schmidt}, \citenamefont {Pradhan},\ and\ \citenamefont
  {Lipson}}]{Xu2005}%
  \BibitemOpen
  \bibfield  {author} {\bibinfo {author} {\bibfnamefont {Q.}~\bibnamefont
  {Xu}}, \bibinfo {author} {\bibfnamefont {B.}~\bibnamefont {Schmidt}},
  \bibinfo {author} {\bibfnamefont {S.}~\bibnamefont {Pradhan}}, \ and\
  \bibinfo {author} {\bibfnamefont {M.}~\bibnamefont {Lipson}},\ }\href@noop {}
  {\bibfield  {journal} {\bibinfo  {journal} {Nature}\ }\textbf {\bibinfo
  {volume} {435}},\ \bibinfo {pages} {325} (\bibinfo {year}
  {2005})}\BibitemShut {NoStop}%
\bibitem [{\citenamefont {Janner}\ \emph {et~al.}(2009)\citenamefont {Janner},
  \citenamefont {Tulli}, \citenamefont {Garc{\'{i}}a-Granda}, \citenamefont
  {Belmonte},\ and\ \citenamefont {Pruneri}}]{Janner2009}%
  \BibitemOpen
  \bibfield  {author} {\bibinfo {author} {\bibfnamefont {D.}~\bibnamefont
  {Janner}}, \bibinfo {author} {\bibfnamefont {D.}~\bibnamefont {Tulli}},
  \bibinfo {author} {\bibfnamefont {M.}~\bibnamefont {Garc{\'{i}}a-Granda}},
  \bibinfo {author} {\bibfnamefont {M.}~\bibnamefont {Belmonte}}, \ and\
  \bibinfo {author} {\bibfnamefont {V.}~\bibnamefont {Pruneri}},\ }\href@noop
  {} {\bibfield  {journal} {\bibinfo  {journal} {Laser and Photonics Reviews}\
  }\textbf {\bibinfo {volume} {3}},\ \bibinfo {pages} {301} (\bibinfo {year}
  {2009})}\BibitemShut {NoStop}%
\bibitem [{\citenamefont {Reed}\ \emph {et~al.}(2010)\citenamefont {Reed},
  \citenamefont {Mashanovich}, \citenamefont {Gardes},\ and\ \citenamefont
  {Thomson}}]{Reed2010}%
  \BibitemOpen
  \bibfield  {author} {\bibinfo {author} {\bibfnamefont {G.~T.}\ \bibnamefont
  {Reed}}, \bibinfo {author} {\bibfnamefont {G.}~\bibnamefont {Mashanovich}},
  \bibinfo {author} {\bibfnamefont {F.~Y.}\ \bibnamefont {Gardes}}, \ and\
  \bibinfo {author} {\bibfnamefont {D.~J.}\ \bibnamefont {Thomson}},\
  }\href@noop {} {\bibfield  {journal} {\bibinfo  {journal} {Nat. Photonics}\
  }\textbf {\bibinfo {volume} {4}},\ \bibinfo {pages} {518} (\bibinfo {year}
  {2010})}\BibitemShut {NoStop}%
\bibitem [{\citenamefont {Tien}\ \emph {et~al.}(2011)\citenamefont {Tien},
  \citenamefont {Bauters}, \citenamefont {Heck}, \citenamefont {Spencer},
  \citenamefont {Blumenthal},\ and\ \citenamefont {Bowers}}]{Tien2011}%
  \BibitemOpen
  \bibfield  {author} {\bibinfo {author} {\bibfnamefont {M.-C.}\ \bibnamefont
  {Tien}}, \bibinfo {author} {\bibfnamefont {J.~F.}\ \bibnamefont {Bauters}},
  \bibinfo {author} {\bibfnamefont {M.~J.~R.}\ \bibnamefont {Heck}}, \bibinfo
  {author} {\bibfnamefont {D.~T.}\ \bibnamefont {Spencer}}, \bibinfo {author}
  {\bibfnamefont {D.~J.}\ \bibnamefont {Blumenthal}}, \ and\ \bibinfo {author}
  {\bibfnamefont {J.~E.}\ \bibnamefont {Bowers}},\ }\href@noop {} {\bibfield
  {journal} {\bibinfo  {journal} {Opt. Express}\ }\textbf {\bibinfo {volume}
  {19}},\ \bibinfo {pages} {13551} (\bibinfo {year} {2011})}\BibitemShut
  {NoStop}%
\bibitem [{\citenamefont {Asano}\ \emph {et~al.}(2017)\citenamefont {Asano},
  \citenamefont {Ochi}, \citenamefont {Takahashi}, \citenamefont {Kishimoto},\
  and\ \citenamefont {Noda}}]{Asano2017}%
  \BibitemOpen
  \bibfield  {author} {\bibinfo {author} {\bibfnamefont {T.}~\bibnamefont
  {Asano}}, \bibinfo {author} {\bibfnamefont {Y.}~\bibnamefont {Ochi}},
  \bibinfo {author} {\bibfnamefont {Y.}~\bibnamefont {Takahashi}}, \bibinfo
  {author} {\bibfnamefont {K.}~\bibnamefont {Kishimoto}}, \ and\ \bibinfo
  {author} {\bibfnamefont {S.}~\bibnamefont {Noda}},\ }\href@noop {} {\bibfield
   {journal} {\bibinfo  {journal} {Opt. Express}\ }\textbf {\bibinfo {volume}
  {25}},\ \bibinfo {pages} {1769} (\bibinfo {year} {2017})}\BibitemShut
  {NoStop}%
\bibitem [{\citenamefont {Zhang}\ \emph {et~al.}(2017)\citenamefont {Zhang},
  \citenamefont {Wang}, \citenamefont {Cheng}, \citenamefont {Shams-Ansari},\
  and\ \citenamefont {Lon\v{c}ar}}]{Zhang2017}%
  \BibitemOpen
  \bibfield  {author} {\bibinfo {author} {\bibfnamefont {M.}~\bibnamefont
  {Zhang}}, \bibinfo {author} {\bibfnamefont {C.}~\bibnamefont {Wang}},
  \bibinfo {author} {\bibfnamefont {R.}~\bibnamefont {Cheng}}, \bibinfo
  {author} {\bibfnamefont {A.}~\bibnamefont {Shams-Ansari}}, \ and\ \bibinfo
  {author} {\bibfnamefont {M.}~\bibnamefont {Lon\v{c}ar}},\ }\href@noop {}
  {\bibfield  {journal} {\bibinfo  {journal} {Optica}\ }\textbf {\bibinfo
  {volume} {4}},\ \bibinfo {pages} {1536} (\bibinfo {year} {2017})}\BibitemShut
  {NoStop}%
\bibitem [{\citenamefont {Nguyen}\ \emph {et~al.}(2011)\citenamefont {Nguyen},
  \citenamefont {Sakai}, \citenamefont {Shinkawa}, \citenamefont {Ishikura},\
  and\ \citenamefont {Baba}}]{Nguyen2011}%
  \BibitemOpen
  \bibfield  {author} {\bibinfo {author} {\bibfnamefont {H.~C.}\ \bibnamefont
  {Nguyen}}, \bibinfo {author} {\bibfnamefont {Y.}~\bibnamefont {Sakai}},
  \bibinfo {author} {\bibfnamefont {M.}~\bibnamefont {Shinkawa}}, \bibinfo
  {author} {\bibfnamefont {N.}~\bibnamefont {Ishikura}}, \ and\ \bibinfo
  {author} {\bibfnamefont {T.}~\bibnamefont {Baba}},\ }\href@noop {} {\bibfield
   {journal} {\bibinfo  {journal} {Opt. Express}\ }\textbf {\bibinfo {volume}
  {19}},\ \bibinfo {pages} {13000} (\bibinfo {year} {2011})}\BibitemShut
  {NoStop}%
\bibitem [{\citenamefont {Lira}\ \emph {et~al.}(2012)\citenamefont {Lira},
  \citenamefont {Yu}, \citenamefont {Fan},\ and\ \citenamefont
  {Lipson}}]{Lira2012}%
  \BibitemOpen
  \bibfield  {author} {\bibinfo {author} {\bibfnamefont {H.}~\bibnamefont
  {Lira}}, \bibinfo {author} {\bibfnamefont {Z.}~\bibnamefont {Yu}}, \bibinfo
  {author} {\bibfnamefont {S.}~\bibnamefont {Fan}}, \ and\ \bibinfo {author}
  {\bibfnamefont {M.}~\bibnamefont {Lipson}},\ }\href@noop {} {\bibfield
  {journal} {\bibinfo  {journal} {Phys. Rev. Lett.}\ }\textbf {\bibinfo
  {volume} {109}},\ \bibinfo {pages} {033901} (\bibinfo {year}
  {2012})}\BibitemShut {NoStop}%
\bibitem [{\citenamefont {Wang}\ \emph {et~al.}(2017)\citenamefont {Wang},
  \citenamefont {Zhang}, \citenamefont {Stern}, \citenamefont {Lipson},\ and\
  \citenamefont {Loncar}}]{Wang2017}%
  \BibitemOpen
  \bibfield  {author} {\bibinfo {author} {\bibfnamefont {C.}~\bibnamefont
  {Wang}}, \bibinfo {author} {\bibfnamefont {M.}~\bibnamefont {Zhang}},
  \bibinfo {author} {\bibfnamefont {B.}~\bibnamefont {Stern}}, \bibinfo
  {author} {\bibfnamefont {M.}~\bibnamefont {Lipson}}, \ and\ \bibinfo {author}
  {\bibfnamefont {M.}~\bibnamefont {Loncar}},\ }\href@noop {} {\ \textbf
  {\bibinfo {volume} {26}},\ \bibinfo {pages} {1547} (\bibinfo {year}
  {2017})}\BibitemShut {NoStop}%
\bibitem [{\citenamefont {Jotzu}\ \emph {et~al.}(2014)\citenamefont {Jotzu},
  \citenamefont {Messer}, \citenamefont {Desbuquois}, \citenamefont {Lebrat},
  \citenamefont {Uehlinger}, \citenamefont {Greif},\ and\ \citenamefont
  {Esslinger}}]{Jotzu2014}%
  \BibitemOpen
  \bibfield  {author} {\bibinfo {author} {\bibfnamefont {G.}~\bibnamefont
  {Jotzu}}, \bibinfo {author} {\bibfnamefont {M.}~\bibnamefont {Messer}},
  \bibinfo {author} {\bibfnamefont {R.}~\bibnamefont {Desbuquois}}, \bibinfo
  {author} {\bibfnamefont {M.}~\bibnamefont {Lebrat}}, \bibinfo {author}
  {\bibfnamefont {T.}~\bibnamefont {Uehlinger}}, \bibinfo {author}
  {\bibfnamefont {D.}~\bibnamefont {Greif}}, \ and\ \bibinfo {author}
  {\bibfnamefont {T.}~\bibnamefont {Esslinger}},\ }\href@noop {} {\bibfield
  {journal} {\bibinfo  {journal} {Nature}\ }\textbf {\bibinfo {volume} {515}},\
  \bibinfo {pages} {237} (\bibinfo {year} {2014})}\BibitemShut {NoStop}%
\bibitem [{\citenamefont {Khanikaev}\ \emph {et~al.}(2015)\citenamefont
  {Khanikaev}, \citenamefont {Fleury}, \citenamefont {Mousavi},\ and\
  \citenamefont {Al{\`u}}}]{Khanikaev2015}%
  \BibitemOpen
  \bibfield  {author} {\bibinfo {author} {\bibfnamefont {A.~B.}\ \bibnamefont
  {Khanikaev}}, \bibinfo {author} {\bibfnamefont {R.}~\bibnamefont {Fleury}},
  \bibinfo {author} {\bibfnamefont {S.~H.}\ \bibnamefont {Mousavi}}, \ and\
  \bibinfo {author} {\bibfnamefont {A.}~\bibnamefont {Al{\`u}}},\ }\href@noop
  {} {\bibfield  {journal} {\bibinfo  {journal} {Nat. Comm.}\ }\textbf
  {\bibinfo {volume} {6}},\ \bibinfo {pages} {8260} (\bibinfo {year}
  {2015})}\BibitemShut {NoStop}%
\end{thebibliography}

%

\end{document}